\documentclass[iop, twocolumn, appendixfloats]{emulateapj}
\usepackage{apjfonts}
\usepackage{natbib}
\usepackage{amsfonts}
\usepackage{amsmath}
\usepackage[usenames]{color}
\usepackage{ulem}
\usepackage{verbatim}
\usepackage{longtable}
\usepackage{array}
\usepackage{epsfig}
\usepackage{scrextend}
\usepackage{mathrsfs}
\usepackage{rotating}
\usepackage{url}
\newcommand\msun{{M_\odot}}

\newcommand\logage{\rm \log(Age)\;[yr]}
\newcommand\teff{T_{\rm eff}}
\newcommand\logg{\log g}
\newcommand\numax{\nu_{\rm max}}
\newcommand\Dnu{\Delta \nu}

\newcommand\amlt{\alpha_{\rm MLT}}

\newcommand\ionn[2]{#1$\;${\scshape{#2}}}
     
\shorttitle{}
\shortauthors{}

\begin{document}

\title{Star Cluster Ages in the {\it Gaia} Era}

\author{Jieun Choi\altaffilmark{1}}
\author{Charlie Conroy\altaffilmark{1}}
\author{Yuan-Sen Ting\altaffilmark{2,3,4}}
\author{Phillip A. Cargile\altaffilmark{1}}
\author{Aaron Dotter\altaffilmark{1}}
\author{Benjamin D. Johnson\altaffilmark{1}}

\altaffiltext{1}{Harvard-Smithsonian Center for Astrophysics, Cambridge, MA 02138, USA}
\altaffiltext{2}{Institute for Advanced Study, Princeton, NJ 08540, USA}
\altaffiltext{3}{Department of Astrophysical Sciences, Princeton University, Princeton, NJ 08544, USA}
\altaffiltext{4}{Observatories of the Carnegie Institution of Washington, Pasadena, CA 91101, USA}

\begin{abstract}
{We use the framework developed as part of the MESA Isochrones and Stellar Tracks (MIST) project to assess the utility of several types of observables in jointly measuring the age and 1D stellar model parameters in star clusters. We begin with a pedagogical overview summarizing the effects of stellar model parameters, such as the helium abundance, mass-loss efficiency, and the mixing length parameter, on observational diagnostics including the color-magnitude diagram, mass-radius relation, and surface abundances, amongst others. We find that these parameters and the stellar age influence observables in qualitatively distinctive, degeneracy-breaking ways. To assess the current state of affairs, we use the recent Gaia Data Release 2 (DR2) along with data from the literature to investigate three well-studied old open clusters---NGC6819, M67, NGC6791---as case studies. Although there is no obvious tension between the existing observations and the MIST models for NGC6819, there are interesting discrepancies in the cases of M67 and NGC6791. At this time, parallax zero point uncertainties in Gaia DR2 remain one of the limiting factors in the analysis of these clusters. With a combination of exquisite photometry, parallax distances, and cluster memberships from Gaia at the end of its mission, we anticipate precise and accurate ages for these and other star clusters in the Galaxy.}
\end{abstract}

\maketitle
\section{Introduction}
\label{section:introduction}

Over the last decade, improving the state of stellar models has become a critical and necessary step in the quest to understand the properties of thousands of exoplanets that have been discovered \citep[e.g.,][]{Torres2012, Huber2014, Mathur2017}, probe the formation and evolution histories of galaxies both near and far including our own Milky Way \citep[e.g.,][]{Freeman2002, Bovy2012, Martig2015}, link the diverse set of transient events to their progenitors \citep[e.g.,][]{Kochanek2008, Smartt2009, Georgy2012}, and interpret the troves of asteroseismology data that have been obtained by the {\it CoRoT} \citep{Baglin2006} and {\it Kepler/K2} missions \citep{Gilliland2010, Bedding2010, Huber2011}. Moreover, it has become increasingly clear that the analysis and interpretation of an even larger wealth of data expected from future missions and surveys will require more complete and accurate stellar models.

Many of the essential ingredients in standard 1D stellar evolution models cannot be modeled from first principles and instead rely on physically-motivated prescriptions. For example, turbulent, superadiabatic convection is usually implemented according to the mixing length formalism in which the mixing efficiency and stellar structure depend sensitively on $\amlt$, a free parameter of order unity \citep{BohmVitense1958}. There are ongoing complementary efforts to address this using sophisticated 3D hydrodynamic simulations \citep[e.g.,][]{Trampedach2014, Magic2015} as well as detailed constraints and calibrations from a variety of observations \citep[e.g.,][]{Bonaca2012, Wu2015, Tayar2017}. This work adopts the latter approach, in particular using well-studied benchmark star clusters with a comprehensive set of observations to investigate the type of information---both cluster and stellar parameters as well as the input physics parameters---that we can recover and the precision with which we can measure them.

This paper is the first in a series that attempts to measure stellar parameters (e.g., age) and constrain uncertain input physics (e.g., mixing length parameter). The insights gained from this work should guide our intuition to both shape the direction of future observations and forecast what we will be able to learn from future surveys and large data sets. We will explore this more quantitatively in subsequent work. In this first paper, we lay the groundwork for our approach by qualitatively examining the effects of various uncertainties on the observable quantities. We explore uncertainties of both observational (e.g., metallicity of the cluster) and theoretical (e.g., efficiency of mass loss) origins \citep[e.g.,][]{Magic2010, Reese2016, Lagarde2017, Angelou2017}. A key aspect of this particular work is that {\it we consider a diverse set of observables simultaneously}. One of the goals is to explore the separation of the key parameters in the various observed planes and identify a set of suitable observables for each parameter.

The rest of the paper is organized as follows. In Section~\ref{section:observations}, we review the different types of data sets that can be employed to study the properties of star clusters and to improve stellar evolution models. In Section~\ref{section:theory}, we first provide a brief overview of the MIST project that serves as the framework for the evolutionary models explored in this work. Then we explore the information content in these observables using theoretical models, paying particular attention to the observational feasibility as well as degeneracies. Next, in Section~\ref{section:case_studies}, we present case studies of three well-studied open clusters, NGC6819, M67, and NGC6791. In Section~\ref{section:gaia}, we discuss what we can expect to accomplish with the future {\it Gaia} data, and in Section~\ref{section:summary}, we present the summary of this work. In the Appendix, we present a series of figures illustrating the effect of uncertain model parameters on the CMD morphologies. For this work, we adopt a \cite{Kroupa2001} initial mass function (IMF) where necessary.

\section{Observations: What Can They Tell Us?}
\label{section:observations}
Here we provide a broad overview of the different types of data sets and surveys, including those that are ongoing and imminent, that can be used to improve both the characterization of star clusters and the quality of the stellar models. We also discuss what type of information can be leveraged from different types of observations. We conclude each section with a discussion of the ``typical'' uncertainties.

\subsection{Photometry}
\label{section:observations_phot}
High-precision photometry in multiple filters covering a long wavelength baseline is tremendously useful for measuring the age, metallicity, extinction, and distance. Stellar evolutionary tracks must be paired with bolometric correction tables to transform the theoretical outputs, e.g., $\teff$ and $\log L$, to observed magnitudes. Under the assumption of perfect observational data, any mismatch between the models and observations can be attributed to one or both of the components: interior models and atmosphere models.

In addition to traditional CMD fitting, photometry can be used for other observational diagnostics such as number counts of different types of stars. Although taking inventory of stars can be a difficult task due to completeness issues as well as low number statistics in some cases, number ratios are still powerful diagnostics because they are sensitive to relative phase lifetimes. We expect to be able to reliably catalog stars in different parts of the CMD with clean membership identification from future surveys (see Section~\ref{section:gaia} for a more in-depth discussion on the improvements due to {\it Gaia}). A related observable is the luminosity function, e.g., along the red giant branch (RGB), which has been widely adopted in studies of globular clusters \citep[e.g.,][]{Renzini1988}.

We also note that multi-band photometry can be used to obtain photometric metallicities (narrow- and medium-band imaging; see e.g., \citealt{Ross2014}) and temperatures. Temperatures derived from color-temperature relations \citep[e.g.,][]{Alonso1996, Ramirez2005, GonzalezHernandez2009, Casagrande2010} are widely used because they are considered to be reliable and easily measurable en masse. Finally, we note that direct measurements of the stellar angular diameter (and physical diameters if the parallax distance is known) for a sample of nearby stars are available through interferometry \citep[e.g.,][]{Boyajian2012a, Boyajian2012b}. In particular, when combined with bolometric flux and multi-band photometry, they provide direct constraints on the empirical color-temperature relations with a few \% accuracy \citep{Boyajian2013}.

Ground-based photometry, which is generally limited by seeing due to the Earth's atmosphere, produces typical uncertainties of order $\approx0.01$ mag, while {\it HST} photometry can routinely yield $\approx$~mmag photometry (relative, not absolute, uncertainty). One source of uncertainty that impacts both ground- and space-based observations is the photometric zero point, which is necessary to convert a flux to a magnitude on some standard scale. Although high-quality photometry can produce high relative photometric precision, absolute photometric precision is tied to $\approx1~\%$ absolute flux uncertainty for flux standards such as Vega (see \citealt{Bohlin2014} and also the discussion in \citealt{Carrasco2016}). Due to the uncertainties associated with the detailed bandpass shape, absolute flux calibration uncertainties for broadband photometry may be as large as $\approx2\%$, and even larger for medium- and narrow-band photometry, for stars with spectral types much different from that of the photometric standards (see Section 3 of \citealt{Bohlin2012} and \citealt{Evans2018}).

\newpage 
\subsection{Spectroscopy}
\subsubsection{Basic Stellar Parameters}
\label{section:spec_basic_stellar_params}
There are several recent, ongoing, and planned large-scale surveys designed to obtain medium-resolution ($R\simeq10,000\textrm{--}25,000$) spectra of stars in different parts of the Milky Way (e.g., {\it RAVE}, \citealt{Steinmetz2006}; {\it Gaia-ESO}, \citealt{Gilmore2012}; {\it APOGEE}, \citealt{Holtzman2015}; {\it GALAH}, \citealt{DeSilva2015}; {\it WEAVE}, \citealt{Dalton2012}; {\it Gaia-RVS}, \citealt{RecioBlanco2016}; {\it 4MOST}, \citealt{deJong2016}). Their principle scientific objective is to shed light on the formation and evolution history of our Galaxy. These spectroscopic surveys yield, at minimum, radial velocity, $\logg$, $\teff$, and metallicity, and in many cases the surface abundances of multiple elements for each star. From the stellar evolution and stellar astrophysics perspective, accurate and precise measurements of these parameters are extremely useful for testing the integrity of the stellar evolution models. With the exception of asteroseismology, surface abundances are some of the only probes of the stellar structure and interior conditions. Since the creation and destruction of different species deep within the star can be imprinted on the surface through various mixing processes, the surface abundances of different elements carry immense diagnostic power. Finally, a Hertzsprung-Russell (HR) diagram constructed from $\logg$ and $\teff$, also known as the ``Kiel diagram,'' is a useful, distance-independent diagnostic that can be compared directly with theoretical isochrones. 

There is immense diversity in the analysis techniques and pipelines that are employed to measure stellar parameters for large spectroscopic samples \citep[e.g.,][]{Smiljanic2014, Holtzman2015}---adopted line lists, optimization for the analysis of different stellar types, equivalent width versus full spectral fitting---and they produce systematically discrepant results. There is ongoing effort to mitigate some of these concerns by carrying out detailed comparisons between different state-of-the-art methods \citep[e.g., {\it Gaia-ESO} benchmark stars;][]{Smiljanic2014}. There is typically a range of values for the quoted uncertainties (combined systematic and statistical), depending on the adopted methodology and the stellar spectral types; $\logg$, $\teff$ and [Fe/H] uncertainties are generally $0.1\textrm{--}0.2$~dex, $50\textrm{--}100$~K, and $0.05\textrm{--}0.1$~dex, respectively \citep[e.g.,][]{Smiljanic2014, Holtzman2015}. Systematic uncertainties are generally higher ($0.1\textrm{--}0.2$) for the other elements, though in some cases, the range of reported [Fe/H] values may be comparable as well \citep[e.g., $\sim0.2$~dex for NGC6791:][]{Carraro2006, Gratton2006, Origlia2006, Carretta2007, Brogaard2011, Boesgaard2015, Netopil2016}.

\subsubsection{Carbon and Nitrogen Surface Abundances on the RGB}
\label{section:cn_rgb_explain}
The surface stellar abundances generally do not reflect the initial or even the bulk interior abundances at any given time. Over the lifetime of a star, physical processes such as diffusion and dredge-up can dramatically modify the surface abundances. The magnitude of these effects vary with the mass and metallicity of the star and the elemental species in question. For this reason, the evolution of surface abundances can be used to trace stellar mass, and thus, stellar ages.

For this work, we focus on the surface abundances of RGB stars because they constitute a significant fraction of the sample in these spectroscopic surveys that require bright beacons in distant parts of the Galaxy (see \citealt{Dotter2017} for a discussion of surface versus bulk abundances and their implications on derived stellar ages). One of the parameters that are crucial to galactic archaeology is the stellar age. The classic method of inferring stellar ages using the spectroscopic $\logg\textrm{--}\teff$ diagram is notoriously challenging due to small $\teff$ separations between the nearly-vertical RGB tracks of stars with different initial masses. As a result, small uncertainties in $\teff$ yield large uncertainties in the initial mass, and therefore in age. 

Recently, an alternative method using the ratio of carbon to nitrogen as an age indicator has gained traction \citep[e.g.,][]{Masseron2015, Salaris2015, Martig2016, Ness2016}. When the star leaves the MS, its deepening convective envelope introduces several changes to the surface elemental abundances during what is known as the first dredge-up (FDU). Whereas some species such as iron that were depleted during the MS due to gravitational settling are nearly restored to their initial values, other species such as nitrogen and carbon show a marked change relative to their initial abundances. The latter phenomenon occurs because the convective envelope engulfs the products of hydrogen burning in the deep interior, diluting its original bulk abundances with the CN-processed material. FDU yields an increase in surface $\rm ^{14}N$ and a concordant decrease in surface $\rm ^{12}C$ as dictated by the CNO cycle equilibrium; the $\rm ^{14}N(p,\gamma)^{15}O$ reaction is the bottleneck in the CNO cycle, resulting in the accumulation of nitrogen. Since the maximum fractional depth reached by the convective envelope increases with the initial mass, a larger decrease in the surface [C/N] abundance corresponds to a larger stellar mass, and therefore a younger age \citep{Salaris2015}. This FDU efficiency has also been demonstrated to increase with increasing metallicity \citep{Charbonnel1994}. A caveat of this age inference method is that the initial abundances must be known (e.g., see the discussion in \citealt{Martig2016}) by disentangling the effects of stellar evolution and galactic chemical evolution. A significant advantage of studying stars in clusters is that we can obtain the abundances for a sample of MS or subgiant branch (SGB) stars in addition to the RGB stars to get a handle on their initial C and N abundances.

The end of the FDU is marked by the convective envelope receding back towards the surface ahead of the hydrogen-burning shell, which is also moving outward. Although canonical models do not show additional mixing beyond the FDU, there is solid observational evidence that extra mixing occurs beyond the RGB bump through the tip of the RGB, and possibly during the core helium burning (CHeB) phase \citep{Gratton2000, Angelou2015}. Several explanations have been put forth, including thermohaline \citep{Charbonnel2007, Charbonnel2010} and rotational mixing \citep{Sweigart1979, Palacios2006}, though we focus on the former here. 

Thermohaline mixing is a double-diffusive instability that occurs in the presence of a destabilizing composition gradient. Although positive mean molecular weight ($\mu$) gradients are rare in the stellar interior (nuclear fusion occurs inside out and it transforms light elements into heavy elements), they do appear in some cases, for example during the $\rm ^{3}He(^{3}He,2p)^{4}He$ reaction taking place in the external wing of the hydrogen-burning shell \citep{Ulrich1972, Eggleton2006}. Note that this unusual reaction produces a net increase in the number of particles and thus a decrease in $\mu$. Thermohaline mixing is established only beyond the RGB bump, a brief adjustment period the star undergoes when the hydrogen-burning shell encounters the $\mu$-discontinuity at the base of the chemically homogeneous zone, i.e., maximum depth previously reached by the convective envelope. This instability cannot be triggered at earlier times because the magnitude of the $\mu$ gradient inversion is negligible in the presence of the stabilizing composition gradient. Once thermohaline mixing is established in the radiative layer between the hydrogen burning shell and the convective envelope, surface abundances of numerous species, including $^{3}$He, $^{12}$C, $^{13}$C, and $^{14}$N, can become modified.

\subsection{Asteroseismology}
Asteroseimology relies on the high-precision monitoring of brightness fluctuations in the light curves originating from stellar oscillations. To date, the {\it CoRoT} \citep{Baglin2006} and {\it Kepler} \citep{Gilliland2010} missions have detected solar-like acoustic oscillations in well over $\approx15,000$ red giants \citep{Kallinger2010a, Stello2013, Huber2014}. The ongoing repurposed {\it K2} mission, the upcoming {\it TESS} mission, and next generation surveys such as {\it WFIRST}, {\it Euclid}, and {\it Plato} are expected to increase the sample size dramatically. 

The detection of oscillations requires taking the Fourier transforms of the time-series photometry. There are two main techniques for the subsequent analysis and physical interpretation of the data. The first method is called ``peakbagging,'' or ``boutique-modeling,'' which involves the detailed modeling of individual peaks in the frequency spectrum. This is a challenging and time-consuming problem due to the sheer number of detected modes as well as the presence of mixed dipole modes \citep{Corsaro2015, Handberg2017}.

The second method, which is more widespread given its simplicity, involves reducing the information in the frequency spectrum to two global parameters: the frequency of maximum power, $\numax$, and the average large frequency separation, $\Dnu$. These parameters can be related to stellar parameters via simple scaling relations:

\begin{eqnarray}
\Dnu &\simeq& \sqrt{\frac{M/M_{\odot}}{(R/R_{\odot})^3}}\Delta \nu_{\odot} \\
\nu_{\rm max} &\simeq& \frac{M/M_{\odot}}{(R/R_{\odot})^2\sqrt{T_{\rm eff}/T_{\rm eff,\odot}}} \nu_{\rm max, \odot} \;,
\end{eqnarray}

where $\Delta \nu_{\odot}=135.1~\rm \mu Hz$, $\nu_{\rm max, \odot}=3100~\rm \mu Hz$, and $T_{\rm eff,\odot}=5777~\rm K$ correspond to the solar values. As can be gleaned from the equations above, $\Dnu$ and $\numax$ are each sensitive to the average density and surface gravity of the star, respectively. Once they are measured from the observed frequency spectrum, $\Dnu$ and $\numax$ can be combined with an external estimate of $\teff$---either from spectroscopy or a color-$\teff$ relation---to yield masses and radii. Alternative forms of the scaling relations exist for when independent estimates of e.g., bolometric luminosity, are available. These scaling relations are used to derive masses and radii of RGB stars in the field en masse \citep{Stello2008, Kallinger2010b, Miglio2012}. Though extremely useful, these simple scaling relations have been demonstrated to harbor systematics and thus various corrections have been proposed to improve their accuracy \citep[e.g.,][]{White2011, Miglio2012, Sharma2016, Viani2017}. There is an ongoing effort to test and validate the accuracy of scaling relations using other independent techniques such as eclipsing binaries \citep[e.g,][]{Gaulme2016}. The general consensus is that the scaling relations provide RGB and RC mass and radius estimates to within $\sim10\%$ and $\sim5\%$, respectively (see \citealt{Viani2017} and references therein). Estimates of the $\logg$ from the scaling relation are better determined; systematic uncertainties and biases are estimated to be around 0.01~dex \citep{Hekker2013, Viani2017}.\footnote{Currently there are interesting, unresolved discrepancies between spectroscopic and asteroseismic $\logg$ for red giants with the {\it SDSS APOGEE} spectroscopy and {\it Kepler} asteroseismology (http://www.sdss.org/dr14/irspec/aspcap/\#calibration). Not only does there appear to a mild metallicity dependence to this discrepancy, but the size of the offset appears to be different for CHeB and red giant branch stars for reasons that are not well-understood. See also \cite{Ting2018}.}

\subsection{Detached Eclipsing Binaries}
Eclipsing binary systems yield stellar masses and radii with exquisite precision, routinely below $\sim3\%$ \citep{Torres2010}. In particular, systems that are well-detached such that both members are effectively undergoing single-star evolution---also known as detached eclipsing binaries (DEBs)---provide a unique opportunity for rigorous tests of stellar evolution models; the masses, radii, and/or luminosities of both binary components must agree within the observational uncertainties at a single age \citep{Andersen1991}. Moreover, DEBs can be used to measure stellar ages without the knowledge of distance and interstellar reddening if they are near the MSTO. DEBs in star clusters are especially useful because they can be combined with CMDs to provide complementary constraints on the models \citep[e.g.,][]{Stetson2003, Brogaard2012, Yakut2015, Brewer2016, Goekay2013}.

Ground-based discoveries and analyses of DEBs trace back well over a century \citep{Russell1912} and have yielded parameters of many stellar systems \citep[e.g.,][]{Popper1967, Harmanec1988, Andersen1991, Torres2010}. The unprecedented, precise monitoring by {\it Kepler} has observed close to 2000 eclipsing binaries, approximately 1400 of which are classified as DEBs \citep{Kirk2016}. Ongoing and future missions such as {\it Gaia}, {\it TESS} and {\it PLATO} are expected to find many more eclipsing binaries.

Accurate and precise masses and orbital parameters require high-quality radial velocities measured from double-lined spectra with high spectral resolution and signal-to-ratio. In a single-lined system, where only the primary component is detected, a full orbital solution generally cannot be obtained. In these systems, the total mass must be combined with the mass ratio inferred from the light curves to obtain estimates of individual masses. Perhaps unsurprisingly, masses obtained using this method are generally less reliable due to correlations and degeneracies among the orbital parameters and their resulting solutions \citep{Kirk2016}. Light curves provide stellar radii and orbital parameters, which can be compared with the spectroscopic determinations as a consistency check. As noted earlier, DEBs generally provide masses and radii measurements to within $\sim3\%$ \citep{Torres2010}, and even $<1\%$ in some cases \citep[e.g.,][]{Brewer2016}.

\begin{figure*}
\centering
\includegraphics[width=0.77\textwidth]{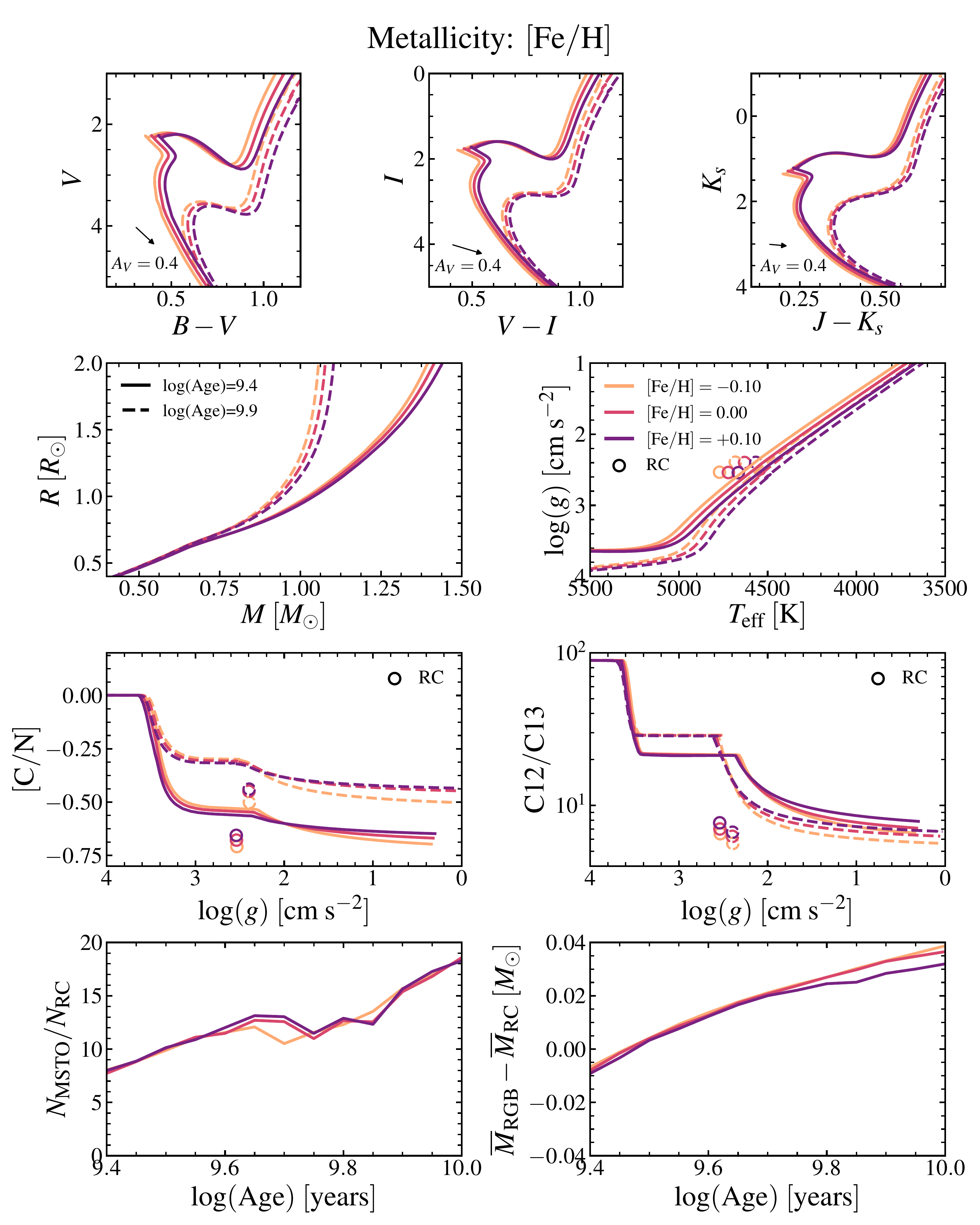}
\caption{The effect of metallicity and age on a variety of observable parameters. There are a total of six MIST models shown, where the peach, pink, and purple colors correspond to different metallicities and the solid and dashed lines correspond to two different ages (2.5 and 8~Gyr). First row: CMDs ranging from optical to near-infrared wavelengths. There are no extinction and distance modulus applied to these CMDs, but we include reddening vectors to illustrate how the positions of the CMDs can shift. We adopt the standard $R_V = A_V/E(B-V)=3.1$ reddening law from \cite{Cardelli1989} and assume $A_V=0.4$ evaluated at $\rm [Fe/H]=0$, $\logg=4$, and $\teff=5500$~K. Second row: mass-radius and Kiel diagrams, which can be compared with DEB data and asteroseismic $\logg$, respectively. For display purposes, we omit the transition from the core He flash (RGB tip) to the RC and mark the start of quiescent core helium burning using an open circle. Third row: the surface abundance evolution as the star ascends the RGB and undergoes helium flash before settling into a quiescent, helium-burning RC phase. The first large decrease at $\logg\sim3.5$ marks the onset of first dredge-up, and the second large decrease at $2.5 \gtrsim \logg \gtrsim 2.0$ after the RGB bump is due to thermohaline mixing. Fourth row: the number ratio of MSTO to RC stars and the difference in the average mass of the RGB and RC stars, both shown as a function of the cluster age. MSTO stars are defined to be those that fall within 0.5~magnitude below the hottest point of the MS in the $V$-band. Asteroseismology can be used to both distinguish RC from RGBs and provide average masses for both types of stars. The first two rows clearly demonstrate that higher metallicity corresponds to cooler $\teff$ (redder colors) and longer MS lifetimes. The third row shows that FDU and thermohaline operate more and less efficiently, respectively, as the metallicity is increased. Finally, the last row shows that metallicity does not appear to have a strong effect on the relative numbers of MSTO to RC stars and the difference in the average mass of RGB and RC stars.}
\label{fig:vary_metallicity}
\end{figure*}

\section{Models: Where is the Information Content?}
\label{section:theory}
In this section, we first provide an overview of the MESA Isochrones and Stellar Tracks (MIST) models. Then we present a summary of the effects of various uncertainties on the observable quantities. Next, we evaluate the sensitivity of the observables to each key parameter in order to identify the ideal set of observations with the goal of measuring the cluster parameters (e.g., age) and constraining the uncertain free parameters (e.g., $\alpha_{\rm MLT}$). We will revisit the latter within a more quantitative and rigorous framework in subsequent work.

\subsection{MESA Isochrones and Stellar Tracks (MIST) Models}
\label{section:mist}
The theoretical isochrones for this work are computed within the MIST framework \citep{Dotter2016, Choi2016}. The main objective of the MIST project\footnote{http://waps.cfa.harvard.edu/MIST/} is to build comprehensive grids of well-calibrated stellar evolutionary tracks and isochrones that encompass a wide range of masses, ages, metallicities, and evolutionary phases. The first set of models with solar-scaled abundances both including and excluding the effects of stellar rotation are already available \citep{Choi2016}. The second set of models consisting of non-solar-scaled abundances are currently under development (Dotter et al., in prep.). Stellar evolutionary tracks are computed using Modules for Experiments in Stellar Astrophysics (MESA) \citep{Paxton2011, Paxton2013, Paxton2015, Paxton2018}, an open-source 1D stellar evolution package. Isochrones are constructed from grids of stellar evolutionary tracks following \citealt{Dotter2016}. For an in-depth overview of the MIST models, including the descriptions of the input physics, we refer the reader to Section~3 and Table~1 of \cite{Choi2016}. 

\subsection{Overview: Model Parameters and Observations}
\begin{figure*}
\centering
\includegraphics[width=0.77\textwidth]{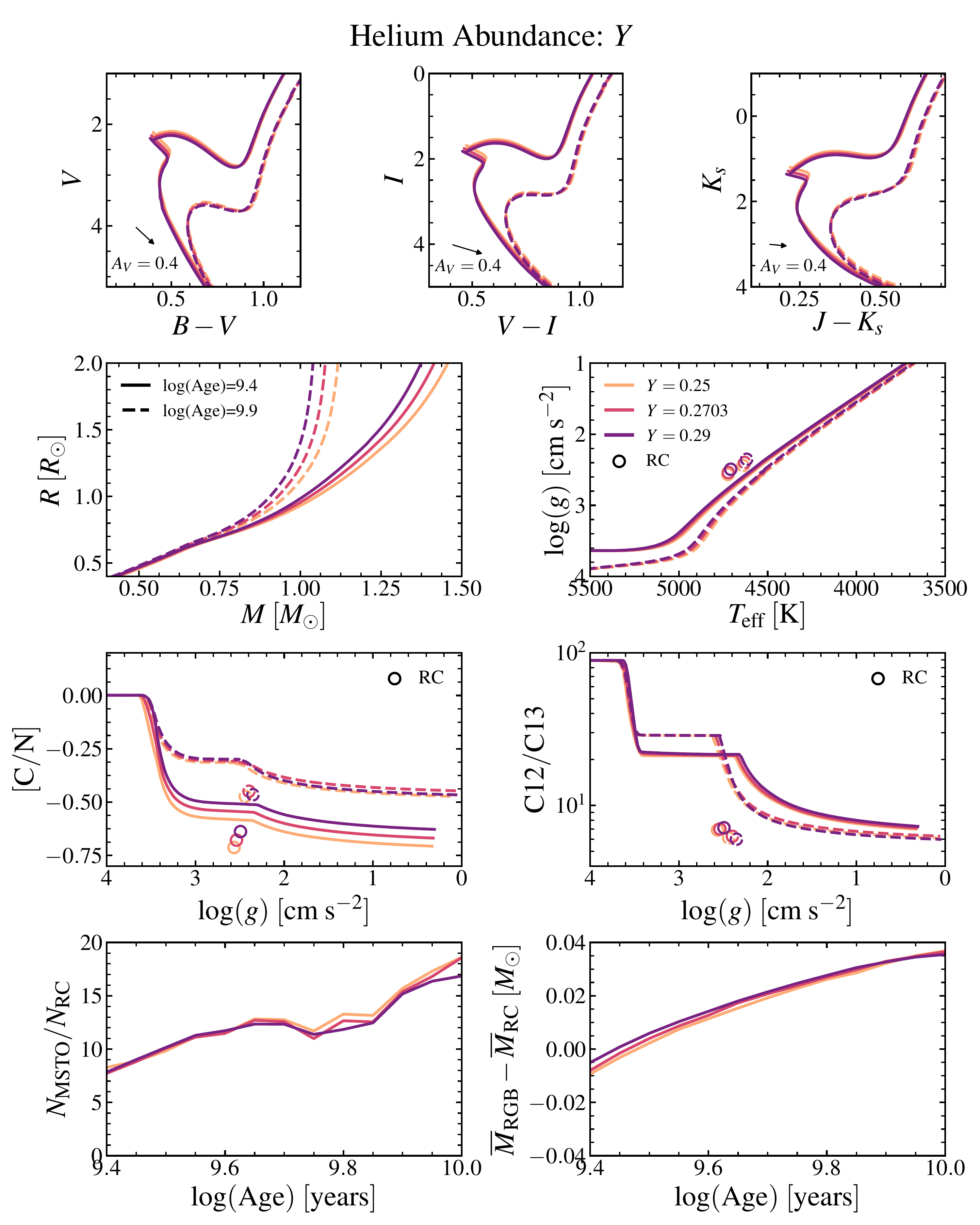}
\caption{Same as Figure~\ref{fig:vary_metallicity} except now varying the helium abundance at a fixed metal mass fraction $Z$. The fiducial value of $Y=0.2703$ comes from the protosolar helium abundance in \cite{Asplund2009}. At a fixed stellar age, higher helium content leads to hotter stars and shorter MS lifetimes.}
\label{fig:vary_helium}
\end{figure*}

\begin{figure*}
\centering
\includegraphics[width=0.77\textwidth]{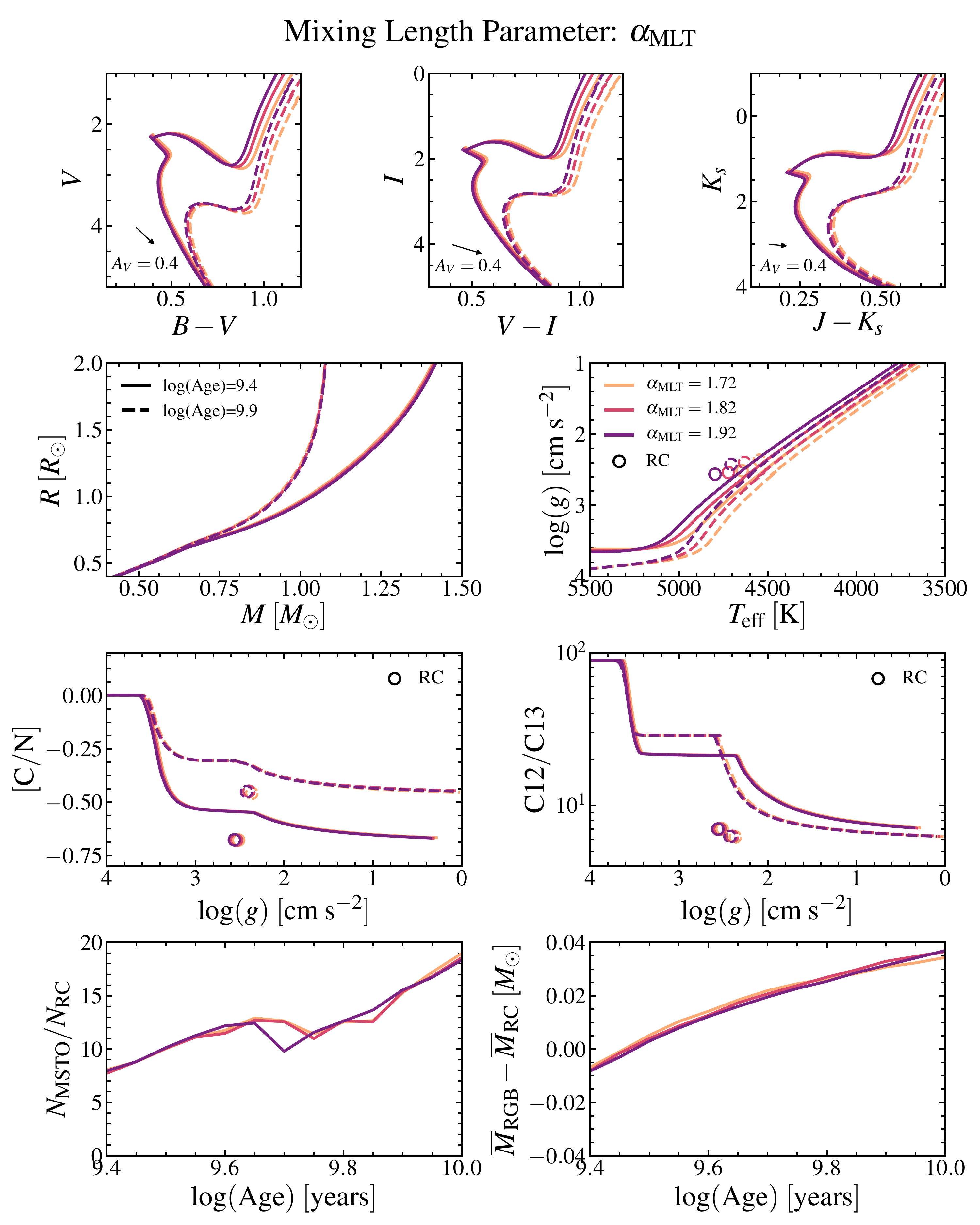}
\caption{Same as Figure~\ref{fig:vary_metallicity} except now varying the mixing length $\amlt$, which parameterizes the efficiency of convective mixing. The fiducial value of $\amlt=1.82$ was chosen by calibrating to the observations of the Sun (see Section 4.1 in \citealt{Choi2016} for more details). The choice of $\amlt$ has the largest effect on the temperature and therefore the color of the RGB.}
\label{fig:vary_amlt}
\end{figure*}

\begin{figure*}
\centering
\includegraphics[width=0.77\textwidth]{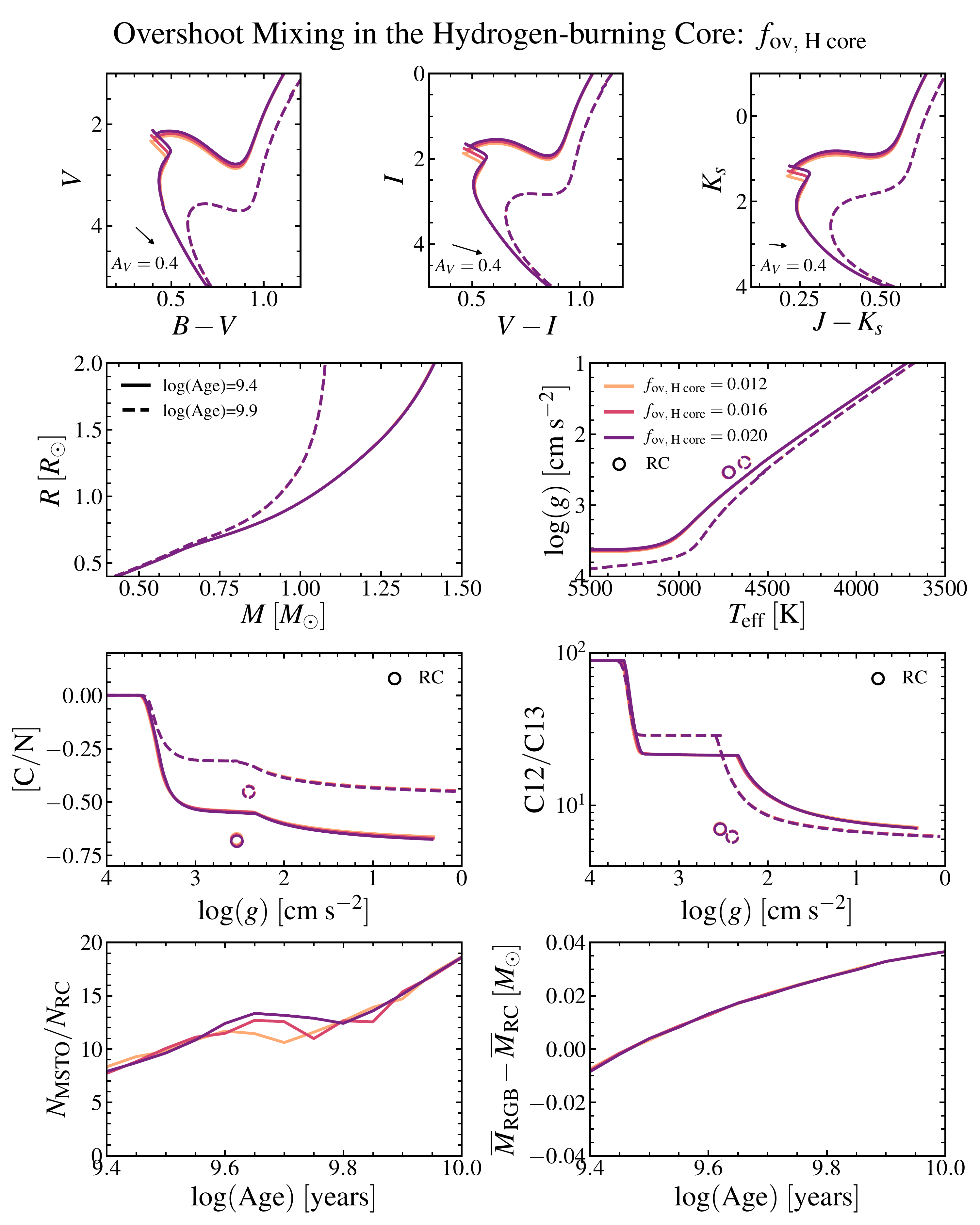}
\caption{Same as Figure~\ref{fig:vary_metallicity} except now varying the efficiency of convective overshoot mixing in the hydrogen-burning core, $f_{\rm ov,\;H\;core}$. The fiducial value of $f_{\rm ov,\;H\;core}=0.016$ was chosen to reproduce the MSTO morphology of the open cluster M67. Independent constraints from DEBs \citep{Claret2016} also lend support for this calibrated value. The choice of $f_{\rm ov,\;H\;core}$ most noticeably affects the MSTO morphology and the luminosity of the SGB because the enhanced mixing during the MS leads to longer MS lifetimes (thus a larger MSTO mass at a fixed age) and a larger core. However, note that this has no effect on an old population because the MSTO stars are low in mass and thus harbor radiative cores.}
\label{fig:vary_fovHcore}
\end{figure*}

\begin{figure*}
\centering
\includegraphics[width=0.77\textwidth]{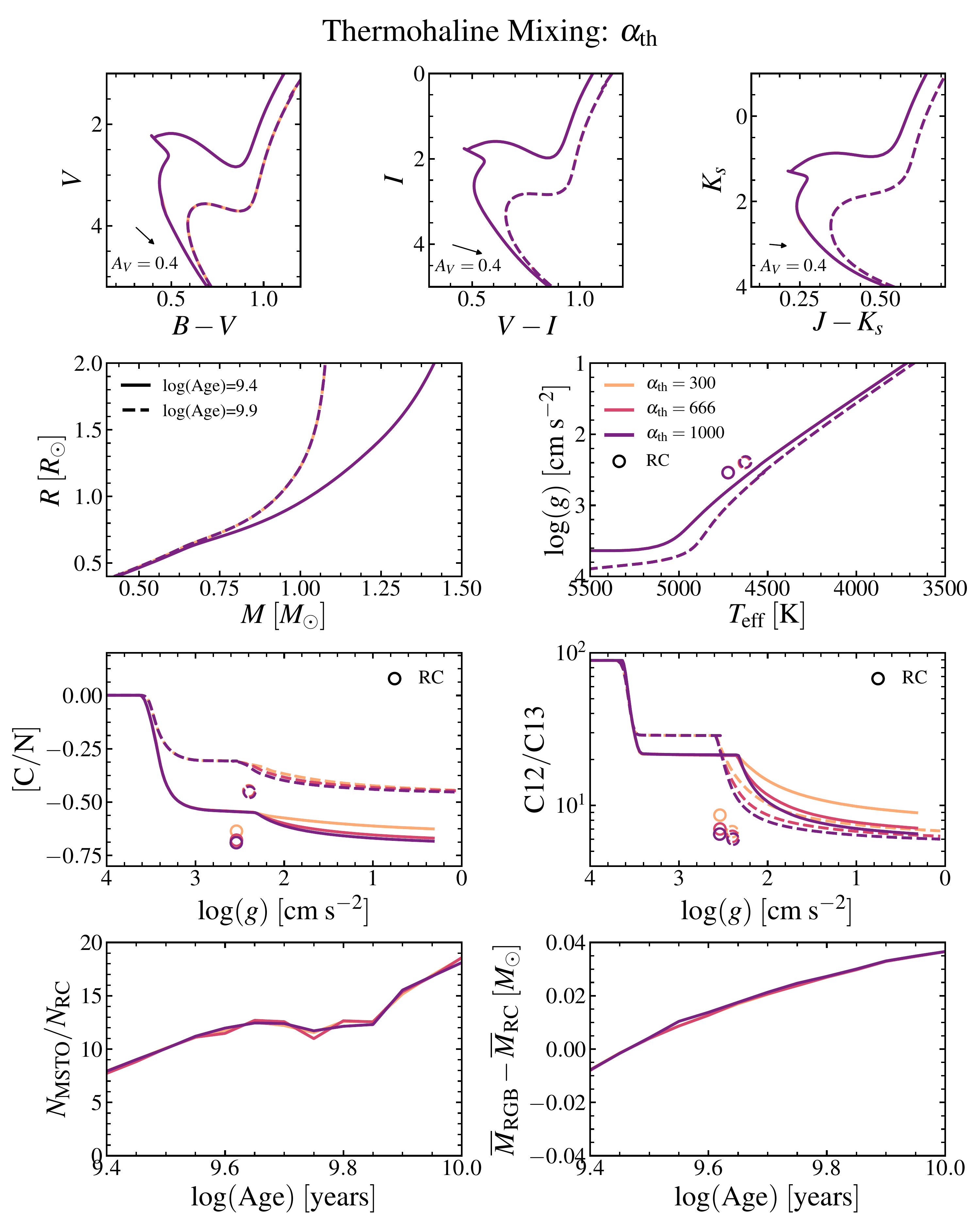}
\caption{Same as Figure~\ref{fig:vary_metallicity} except now varying the efficiency of thermohaline mixing $\alpha_{\rm th}$. The fiducial value of $\alpha_{\rm th}$ was recommended by \cite{Charbonnel2007} to reproduce the observed surface abundances of stars brighter than the RGB bump. The choice of $\alpha_{\rm th}$ has essentially no distinguishable effect on any of the observables except for the surface abundances of RGB stars brighter than the bump.}
\label{fig:vary_thmhln}
\end{figure*}

\begin{figure*}
\centering
\includegraphics[width=0.77\textwidth]{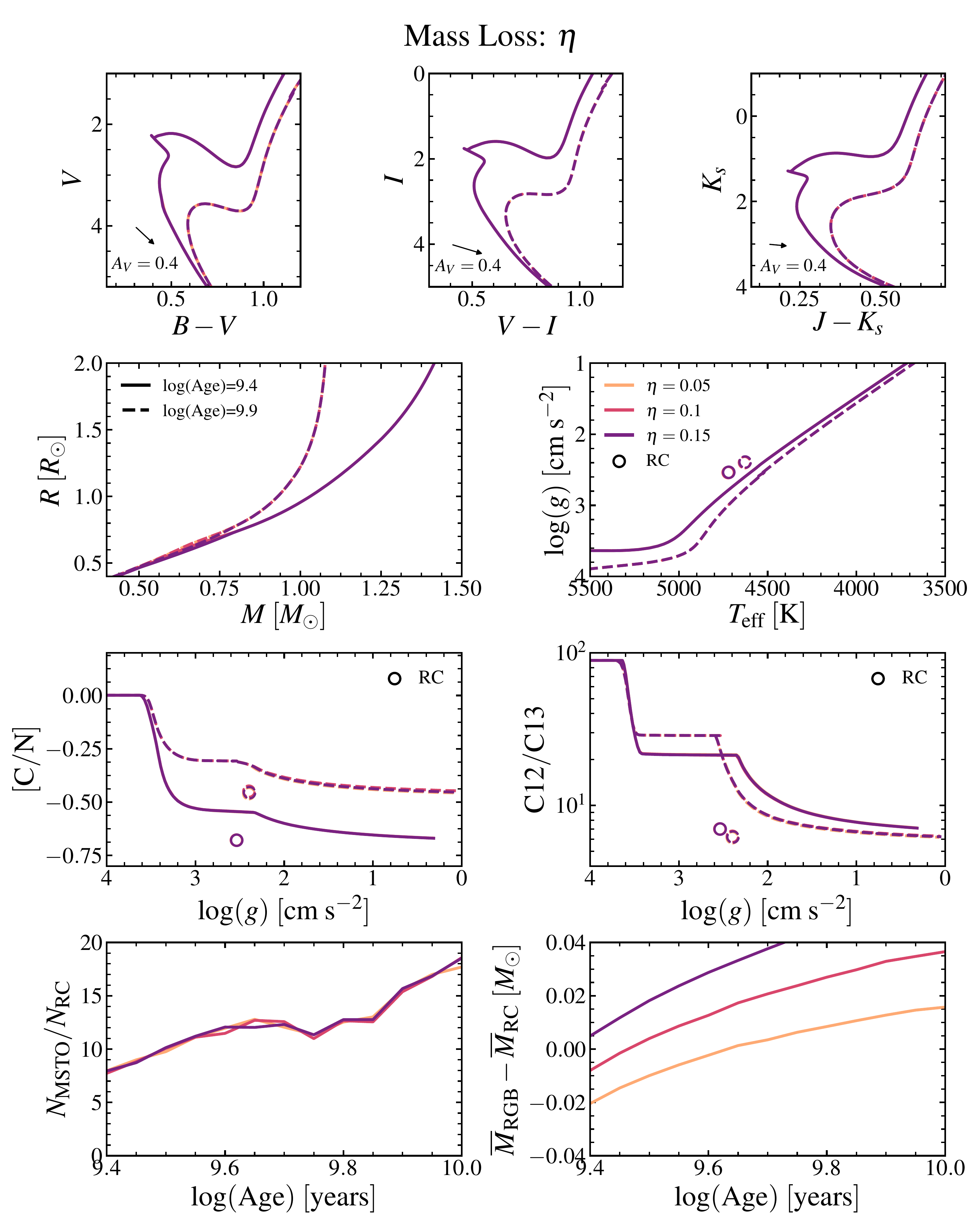}
\caption{Same as Figure~\ref{fig:vary_metallicity} except now varying the Reimers mass loss parameter $\eta$. The fiducial value of $\eta=0.1$ was recommended by the {\it Kepler} asteroseismic constraints from open clusters \citep{Miglio2012, Handberg2017}. It also reproduce the initial-final mass relation (see Section 8.2 in \citealt{Choi2016} for more details). The choice of $\eta$ has almost no discernible effect on any of the observables except for the masses of RGB and RC stars.}
\label{fig:vary_mdoteta}
\end{figure*}

In the following sections, we show a series of multi-panel plots (Figures~\ref{fig:vary_metallicity} through \ref{fig:vary_mdoteta}) illustrating the effect of an individual parameter on the various observables at a given stellar age. Here we describe each panel and the relevant observations in detail.

In each panel, we show a total of nine MIST isochrones projected onto several observed planes. The peach, pink, and purple colors correspond to different parameters (for example, metallicities) and the solid, dot-dashed, and dotted lines correspond to three ages (2, 4, and 10~Gyr). The parameter of interest always increases from peach to purple, and the pink curve corresponds to the fiducial model, unless noted otherwise.
 
The top row features three CMDs---$B-V$, $V-I$, and $J-Ks$---zoomed in near the main sequence turn off (MSTO), SGB, RGB, and RC. As in the standard MIST models, we use the C3K bolometric correction tables (Conroy et al., in prep.) constructed from the ATLAS12/SYNTHE atmosphere models. The ATLAS12/SYNTHE models include the latest atomic line list from R. Kurucz (including both laboratory and predicted lines) and many molecules whose contributions are important especially at longer wavelengths. We use the latest {\it Gaia} DR2 passbands and zero points \citep{Evans2018}. There are no extinction and distance modulus applied to these CMDs.  

In the first panel of the middle row, we show the theoretical mass-radius relations, which can be compared to high-precision measurements from DEBs. The second panel shows a slight variation on the classic HR diagram with $\logg$ instead of $\log L$ on the $y$-axis (Kiel diagram), zoomed in on the RGB and the RC where most of the asteroseismic targets are located. For these evolutionary phases, the predicted $\logg$ can be compared with the asteroseismic surface gravity, $\log g_{\rm astero}$, inferred from the $\numax$ asteroseismic scaling relation (Equation~1). Note that the $\numax$ measurement must be combined with a spectroscopic or photometric $\teff$ to infer $\logg$. For display purposes, we omit the transition from the core helium flash (the tip of the RGB) to the RC and mark the start of quiescent core helium burning using an open circle.

The next row shows surface abundance ratios of carbon to nitrogen (left) and $^{12}$C to $^{13}$C (right) along the RGB and RC (shown as an open circle for clarity) as a function of surface gravity. Surface abundances are powerful indirect probes of the stellar interior because various mixing processes lead to changes in the surface abundances of different species at different stages of the evolution.

Finally, the bottom row features two panels that each shows an integrated or averaged quantity as a function of the age of the cluster. The first panel shows the ratio of MSTO to RC stars. In this context, the MSTO stars are defined to be those that fall within 0.5~magnitude below the hottest point of the MS in the $V$ band. For an isochrone with a Henyey hook, instead of the tip of the hook (the actual hottest point), we use the inflection point at which the blueward excursion begins. This is because the actual hook corresponds to a fast phase of expansion and contraction, and thus it is observationally unlikely to find many stars there. The RC stars are selected based on the MIST phase label (CHeB phase). The predicted number ratio of MSTO and RC stars is simply the ratio of the sum of the IMF weights of stars of each type. This quantity represents the relative phase lifetimes averaged over the IMF. In the right panel, we show the average mass difference between the RGB and the RC stars. Note that this is currently observable stellar mass instead of the initial mass, and that this is a simple average without the IMF weights. We can safely ignore the IMF weights in this case due to the negligible dynamic range in mass. RGB stars are first identified using the phase label in the MIST isochrones, then we apply further selection cuts using the criteria adopted by \cite{Miglio2012}; we discard stars that are brighter than the RC magnitude to reduce possible confusion with asymptotic giant branch (AGB) stars, and fainter than 1.5~mag below the RC in the $V$ band to avoid possible blending and low signal-to-noise issues. RC stars are simply selected by their MIST phase label. The resulting prediction can be compared to the ``observed'' mass difference inferred from asteroseismic masses \citep[e.g.,][]{Miglio2012}.

In each panel we also include reddening vectors to illustrate how the positions of the CMDs can be shifted due to dust. We adopt the standard $R_V = A_V/E(B-V)=3.1$ reddening law from \cite{Cardelli1989} and assume $A_V=0.4$ evaluated at $\rm [Fe/H]=0$, $\logg=4$, and $\teff=5500$. Reddening may appear to be degenerate with metallicity especially at the MSTO, SGB and RGB. But there are subtle differences in the CMD morphology, e.g., the lower MS , especially when multiple CMDs covering a wide wavelength baseline are considered simultaneously (e.g., the ``kink'' in the lower MS in infrared CMDs; see \citealt{Pulone1998, Milone2014, Correnti2016}). Given exquisite photometry (e.g., {\it Hubble Space Telescope} or {\it Gaia}) and sophisticated fitting techniques, we should be able to distinguish the two effects.

\subsection{Effect of Metallicity}
Figure~\ref{fig:vary_metallicity} shows the effect of varying [Fe/H], more specifically $Z/X$, the ratio of metal to hydrogen mass fractions. Note that we assume solar-scaled abundances for the models considered here, i.e., initial [C/N] is held constant. The initial helium mass fraction $Y$ is computed assuming a linear enrichment law, a common approach also adopted in MIST. More specifically, the helium abundance is tied to the metallicity assuming a linear relationship, i.e., $Y=Y_{\rm p} + (Y_{\odot,\rm\;protosolar} - Y_{\rm p})Z/Z_{\odot,\rm\;protosolar}$, where $Y_{\rm p}$ is the primordial, Big Bang nucleosynthesis value. The enrichment slope in MIST is $\Delta Y/\Delta Z=(Y_{\odot,\rm\;protosolar} - Y_{\rm p})/Z_{\odot,\rm\;protosolar}=1.5$. As expected, the CMDs show that increasing metallicity leads to redder colors. The mass-radius panel clearly shows that there is little separation in radius until the stars evolve away from the MS, which suggests that the sensitivity of the models to variations in the age and metallicity is concentrated in the post-MS stars. At a fixed initial mass, metal-rich stars are cooler and have longer MS lifetimes---the Kiel diagram show a clear sequence in temperature with metallicity.

Both of the surface abundances panels show a large dip between the first two plateaus ($\logg \sim3.5$ to $\logg \sim 3$), corresponding to the FDU (see Section~\ref{section:cn_rgb_explain}) and the subsequent decrease is due to thermohaline mixing that is established beyond the RGB bump. Interestingly, the net change in [C/N] and $^{12}$C/$^{13}$C during the two stages of mixing (FDU and thermohaline) show opposite trends with metallicity: FDU and thermohaline mixing appear to operate more efficiently in high and low metallicity systems, respectively. FDU is more efficient at higher metallicities and higher stellar masses because the convective envelope reaches deeper into the CN-processed region and is thus able to dredge up more material to the surface \citep[e.g.,][]{Charbonnel1994, Salaris2015, Lagarde2017}. Thermohaline is more efficient at lower metallicities and lower stellar masses \citep[e.g.,][]{Charbonnel2007, Eggleton2008, Charbonnel2010, Lagarde2017} due to the compactness of the thermohaline-mixing region and the resulting steeper temperature gradient.

The bottom panels show relatively large variation with stellar age but little variation with metallicity, suggesting that these integrated quantities are more sensitive diagnostics of the stellar age than the metallicity. The number ratio is sensitive to the age because the MS lifetime increases dramatically with decreasing initial mass but the CHeB lifetime is roughly constant for stars $\lesssim2~\msun$. The CHeB lifetime is relatively insensitive to the initial mass for these stars because they have roughly equal-sized degenerate helium cores that ignite once a critical temperature (corresponding to a critical mass of $\sim0.45~\msun$) is reached.

\subsection{Effect of Helium}
Figure~\ref{fig:vary_helium} shows the effects of varying the initial bulk helium mass fraction, $Y$, on the various observables. We hold $Z$ fixed at the protosolar value $Z_{\odot,\rm\;protosolar}=0.0142$ and vary $Y$ and the hydrogen abundance $X$. Helium abundance cannot be inferred spectroscopically because there are no photospheric helium lines due to their high excitation potential (\citealt{Asplund2009} but see \citealt{Dupree2013}). However, helioseismology provides an indirect probe of the helium abundance in the Sun, relying on the fact that the adiabatic index changes in the \ionn{He}{ii} ionization zone. This technique has yielded a highly precise estimate of the solar helium abundance ($0.2485\pm0.0034$; \citealt{Basu2004}), which is at an apparent tension with the helium abundance required to reproduce the correct solar luminosity and temperature at the solar age for an interior model \citep{Asplund2009}.

The CMDs in the first row demonstrate a systematic trend with $Y$ as they did with metallicity in Figure~\ref{fig:vary_metallicity}. Note that these panels illustrate the effect of helium abundance on the interior structure and evolution only, largely via changes to the opacity. Helium abundance also influences the stellar spectra and therefore the bolometric corrections, but this effect is unexplored in this work. The mass-radius relation and the $\logg\textrm{--}\teff$ panels also show a sequence in $Y$; at a fixed age, higher helium content leads to hotter stars and lower MSTO masses. The hotter temperature is due to helium's low opacity and the decreased MSTO mass (equivalent to a shorter lifetime) is due to the reduced hydrogen abundance. It is interesting to note that the CMDs do not show a clearly separated sequence in $Y$; the effect of helium variation moves ``along'' the isochrone, such that the effect on the CMD is not drastic. On the other hand, the change in the surface abundances, in particular [C/N], during the FDU is less pronounced at high $Y$. This is because a smaller fraction of the star becomes engulfed, or dredged up, as the stellar mass is decreased. Interestingly, the efficiency of thermohaline mixing appears to be largely unaffected by initial $Y$. Finally, the number ratio of MSTO to RC stars does not show a clear sequence in $Y$, and the mass difference between RGB and $RC$ shows marginal separation in $Y$. However, the difference is much smaller than the current observational uncertainties ($\sim0.01~\msun$) and is unlikely to be detectable in the near future.

\subsection{Effect of Mixing Length Parameter}
In Figure~\ref{fig:vary_amlt}, we show the effects of varying the mixing length parameter $\amlt$, a free parameter of order unity that is frequently calibrated to match the observations of the Sun (in MIST, $\amlt=1.82$; \citealt{Choi2016}). The physical interpretation of $\amlt$ is the distance, in units of the pressure scale height, that a fluid parcel travels before depositing its energy and disintegrating into the background. Thus it parameterizes the efficiency of convective mixing and affects the stellar structure: a small $\amlt$ is associated with cooler $\teff$ and inflated radius. This is clearly illustrated in the CMDs, mass-radius relation, and the Kiel diagram, particularly for the RGB stars which have large convective envelopes. Interestingly, the separation between the models with different values of $\amlt$ is larger in the colors than in $\teff$, especially on the upper RGB. Given the typical observational uncertainties of $\sim50\textrm{--}100$~K in $\teff$ \citep[e.g.,][]{Holtzman2015} and tens of mmag in color\footnote{{\it Hubble Space Telescope}, which nominally represents the best-case scenario today, routinely achieves $\lesssim1~$mmag relative uncertainty.} this suggests that CMDs may be preferable to HR diagrams for constraining $\amlt$. However, the CMD approach requires a reddening correction, though a joint fitting of multiple CMDs may help alleviate the issues with degeneracies. Finally, $\amlt$ appears to have a negligible effect on the RGB surface abundances, the number ratio of MSTO to RC stars, and the average mass difference between the RGB and RC phases.

\subsection{Effect of Convective Overshoot Mixing}
The MLT framework on its own offers an incomplete description of convection and requires a separate prescription for extra mixing that occurs at the convective boundaries. This process, also known as convective overshoot, is meant to capture the nonzero momentum of the fluid parcel approaching the boundary of the convection zone and its subsequent penetration into the radiative region \citep{Unno1957, Bohm1963, Shaviv1973}. Overshoot implies enhanced mixing and thus has several observable consequences, including the properties of AGB and post-AGB stars \citep[e.g.,][]{Herwig2000, Herwig2011}, the MS width \citep{Schaller1992}, and the MSTO morphology in open clusters \citep[e.g.,][]{Magic2010}. 

We adopt the ``exponential diffusive overshoot'' framework introduced by \cite{Freytag1996} and implemented in MESA following \cite{Herwig2000}. This prescription is meant to capture both the exponential decay of the convective velocity field and the dissolution of the fluid parcel as a diffusive process. There are two sets of free parameters available for every convective boundary in MESA: $f_{\rm ov}$ and $f_{\rm 0,ov}$. The first parameter, $f_{\rm ov}$, determines the efficiency of overshoot mixing and describes the velocity scale height in terms of the local pressure scale height. The second parameter, $f_{\rm 0,ov}$, determines the location inside the convection zone at which the diffusion coefficient is calculated. For simplicity, we fix the latter to $f_{\rm 0,ov}=0.008$ (half of the fiducial value for $f_{\rm ov}$) as we vary $f_{\rm ov}$ to investigate the role of the efficiency of overshoot mixing.

Figure~\ref{fig:vary_fovHcore} shows the effect of varying $f_{\rm ov}$ in the hydrogen-burning core. Increasing the efficiency of overshoot mixing in the hydrogen core leads to a more prominent MSTO morphology and a brighter SGB due to an enhanced MS lifetime and a larger core. Note that this parameter has no effect on old populations because their MSTO stars are sufficiently low in mass such that they do not harbor convective cores during the MS. 

We also tested the effect of varying $f_{\rm ov}$ in the helium-burning core. Somewhat surprisingly, changing this parameter seems to have little to no effect on any of the observables, therefore the corresponding figure is not shown. \cite{Montalban2013} computed a series of stellar models and adiabatic frequencies and found a correlation between the average value of the asymptotic period spacing ($\Delta$P) and the size of the helium-burning core. In a more recent work, \cite{Arentoft2017} analyzed a sample of red giants in the open cluster NGC6811 and found that overshoot mixing in the helium-burning core does not appear to have a noticeable effect on the resulting $\Dnu$ and $\numax$ as long as overshoot mixing is included during the MS. However, the authors also found that $\Dnu$ and $\Delta$P together has the potential to constrain the efficiency of overshoot mixing in the helium core and shed light on the still-debated presence of breathing pulses \citep{Castellani1985}. Asteroseismic modeling that probes the detailed interior stellar structure may be required to constrain the efficiency of overshoot in the helium-burning core. 

\newpage
\subsection{Effect of Thermohaline Mixing}
In Figure~\ref{fig:vary_thmhln}, we show the effects of varying $\alpha_{\rm th}$, the efficiency of thermohaline mixing. As described in Section~\ref{section:cn_rgb_explain}, thermohaline is a type of mixing that occurs in a thermally stable medium that has a destabilizing composition gradient. Standard models do not predict any further changes to the surface abundances along the RGB at the conclusion of the FDU, but abundance evolution beyond the RGB bump is indeed observed \citep[e.g.,][]{Gratton2000}. Thermohaline mixing is a viable mechanism for explaining this phenomenon (but see also e.g., \citealt{Denissenkov2010, Traxler2011, Wachlin2014}), wherein an unstable composition gradient is established by the $^3$He reaction taking place in the external wing of the hydrogen burning shell. In the framework of \cite{Ulrich1972} and \cite{Kippenhahn1980}, $\alpha_{\rm th}$ has a geometric interpretation---a large value corresponds to a slender fluid element---which is also directly linked to the mixing timescale and thus, the mixing efficiency. The fiducial value of $\alpha_{\rm th}$ in the MIST models was recommended by \cite{Charbonnel2007} to reproduce the observed surface abundances of stars brighter than the RGB bump.

Perhaps unsurprisingly, changing the efficiency of thermohaline mixing has almost no observable influence except in the surface abundances beyond the RGB bump \citep[see also][]{Lagarde2017}. Note that this effect saturates beyond some critical value of $\alpha_{\rm th}$ (pink and purple curves), suggesting that there is a maximum efficiency with which thermohaline mixing operates. Given the very minor influence of thermohaline mixing on the overall evolution, we do not expect noticeable differences among the models in the bottom left panel showing the number ratio of MSTO and RC stars. The small variations are largely due to the presence of very weak breathing pulses occurring at the end of the CHeB phase and thus the size of these variations nominally represents the minimum theoretical uncertainty on this quantity.

\begin{figure*}
\centering
\includegraphics[width=0.75\textwidth]{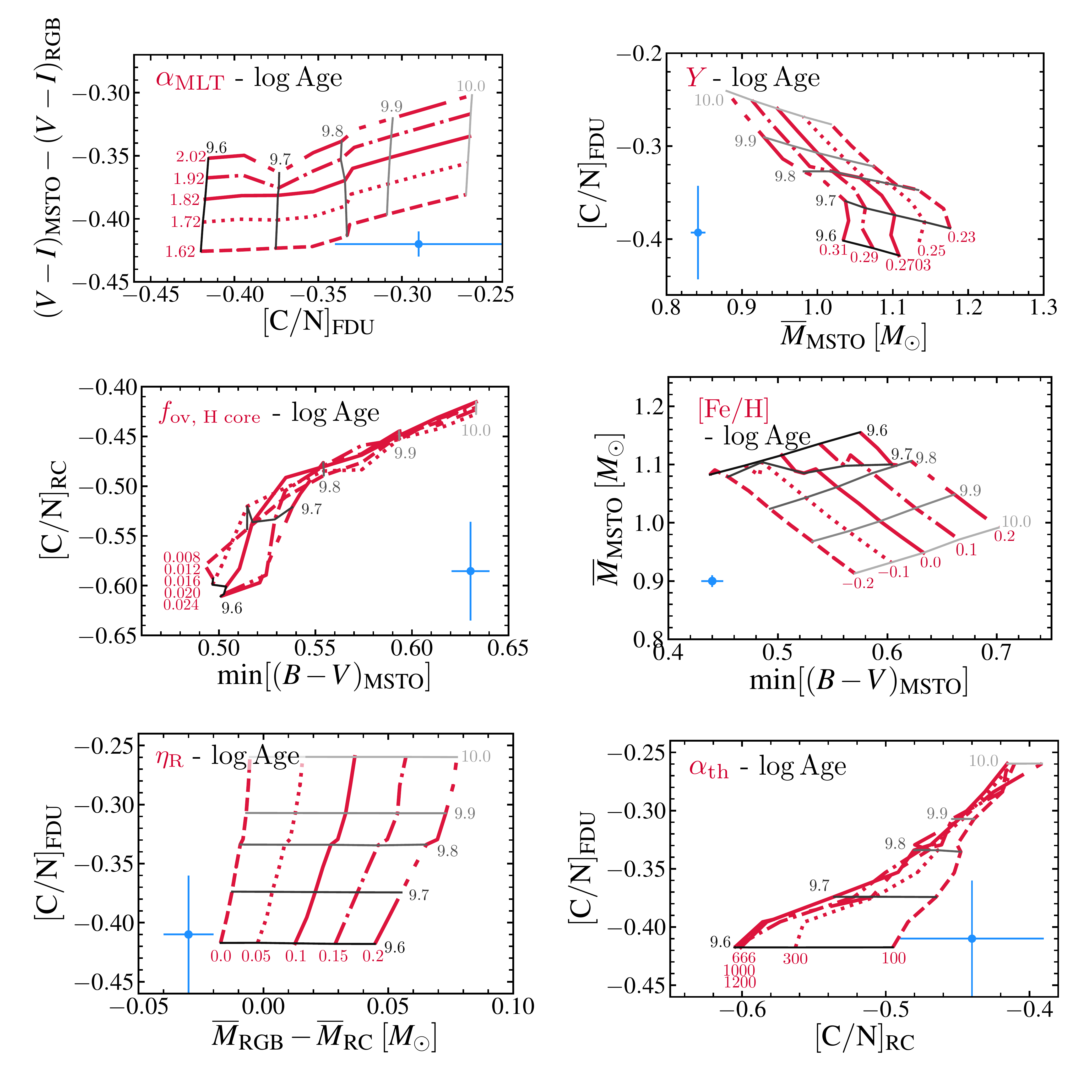}
\caption{Diagnostic sensitivity of a pair of observables to a parameter of interest and the stellar age. For each parameter of interest, we identify a set of observables that most cleanly separate in $\log(\rm Age)$ (gray lines) and the parameter in question (red lines), though this was not always possible in every case. Top left: mixing length parameter; $V-I$ color difference between the MSTO and the RGB (measured at $V=1.5$) vs. surface [C/N] abundance of post-FDU stars below the RGB bump. Top right: helium abundance; surface [C/N] abundance of post-FDU stars below the RGB bump vs. average mass of the MSTO stars. For low values of $Y$, there is no RGB bump at $\log(\rm Age)=9.6$~[years]. Middle left: convective overshoot mixing efficiency in the hydrogen core; surface [C/N] abundance of RC stars vs. $B-V$ color at the MSTO. Middle right: metallicity; average mass of the MSTO stars vs. $B-V$ color at the MSTO. Bottom left: mass loss; surface [C/N] abundance of post-FDU stars below the RGB bump vs. average mass difference between the RGB and RC stars. Bottom right: thermohaline mixing; surface [C/N] abundance of post-FDU stars below the bump vs. surface [C/N] abundance of RC stars. The blue error bar represents a typical observational uncertainty (see Section 2 for an in-depth overview of different observational data sets).}
\label{fig:separation_summary}
\end{figure*}

\subsection{Effect of Mass Loss Efficiency}
Figure~\ref{fig:vary_mdoteta} illustrates the effects of varying the wind efficiency, $\eta$. In particular, since we are focusing on the evolutionary phases preceding the AGB, the relevant mass loss scheme is the \cite{Reimers1975} prescription, where $\dot{M} \propto LR/M$ with a prefactor $\eta$ of order unity. For the fiducial MIST models, we adopt $\eta=0.1$ which is smaller than the value traditionally adopted in stellar models \citep[e.g.,][]{Girardi2000, Pietrinferni2004, Ekstrom2012}. This choice was motivated by the results from \cite{Miglio2012} and more recently \cite{Handberg2017}, who demonstrated that the asteroseismic masses prefer only a modest amount of mass loss on the RGB. The mass loss rate rises steadily as the star ascends the RGB, eventually reaching values as high as $10^{-8}~\msun~\rm yr^{-1}$ at the tip of the RGB. Variations in the mass loss efficiency parameter have almost no effects on the observables considered here except in the average mass difference between the RGB and the RC.\footnote{Although the effect of mass loss on the horizontal branch morphology is not considered in this work, this subject has been studied extensively in the context of globular clusters. See e.g., \citet[][]{Rood1973, Lee1990, Vink2002, Dotter2008b, Catelan2009, Gratton2010, Salaris2016}.} If there is no mass loss on the RGB, this mass difference is always less than zero because the current stellar mass is exactly equal to the initial mass and MS lifetime decreases with increasing stellar mass. On the other hand, if there is significant mass loss between the RGB and the RC, this quantity will be positive. The mass difference is negative-valued at almost all stellar ages for the model with very little mass loss (peach curve), which suggests that evolutionary timescale is the dominant effect in this case.  

\subsection{Other Parameters to Consider}
\label{section:other_parameters_to_consider}
In addition to the ``internal'' sources of uncertainties in stellar models, there are also ``external'' sources of uncertainties due to difficulties associated with measuring abundances, distances, and reddening. Besides metallicity and helium abundance, oxygen abundance (either on its own or grouped with the other $\alpha$-capture elements) is another key parameter in stellar models due to its strong influence on the overall stellar structure and evolution, and thus the inferred stellar age \citep[e.g., ][]{VandenBerg2001, Dotter2007, VandenBerg2012, Bond2013}. Oxygen contributes significantly to the overall opacity in the stellar interior and alters the relative importance of the CNO cycle compared to the pp-chain. We generally assume metallicity\footnote{But not helium!} to be a well-determined quantity from spectroscopy, but as noted in Section~\ref{section:spec_basic_stellar_params}, different methods can yield systematically different measurements at the $\sim0.1~$dex level \citep{Smiljanic2014}. For some species such as oxygen, the spectroscopic abundance may be even more uncertain due to difficulties associated with weak, blended lines, non-local thermodynamic equilibrium effects, and/or 3D effects (see the discussion in \citealt{Asplund2009}, but see also \citealt{Ting2018}). 

Over the next several years, the {\it Gaia} mission will effectively eliminate distance and membership uncertainties for open clusters provided that they are sufficiently nearby. The predicted end-of-mission parallax errors for an individual G2V star at $V=15$ is $24~\mu \rm as$,\footnote{https://www.cosmos.esa.int/web/gaia/science-performance} which corresponds to 2\% and 12\% precision at 1 and 5~kpc, respectively. Even though many clusters, including NGC6791, have MSTOs that are significantly fainter than $V=15$, we should still be able to obtain robust distance estimates by relying on the bright red giants and combining the constraints from many more fainter stars. For reference, the three open clusters considered for this work in Section~\ref{section:case_studies} are at distances of $\sim800$~pc to $\sim4$~kpc, and have MSTO magnitudes of $G=13$ to $17.5$.

Extinction (particularly differential extinction) remains a challenging problem. For this work, we assume the commonly-adopted CCM reddening law \citep{Cardelli1989}, but there are several alternatives including the \citealt{Fitzpatrick1999} and \citealt{ODonnell1994} reddening laws. In practice, $R_V$, which parameterizes the slope of the optical extinction curve, is almost always assumed to be the galactic average $R_V=3.1$ even though there is evidence for variations along different sightlines \citep{Draine2003}. The uncertainties associated with the treatment of extinction may well be a dominant source of uncertainty in our interpretation of CMDs in the {\it Gaia} era. One path forward may be the use of panchromatic CMDs to infer the bolometric magnitude from modeling the spectral energy distribution, which would remove extinction from at least the $y$-axis of the CMD.

\subsection{Separation of Information}
Here we provide a succinct, visual summary of the information presented in the previous sections. Each panel in Figure~\ref{fig:separation_summary} illustrates the sensitivity of a pair of observables to a parameter of interest and the stellar age for the purposes of disentangling their effects. The blue error bar represents a typical observational uncertainty. For some parameters such as $\eta_{\rm R}$ and $\amlt$, their effects are nearly orthogonal to that of stellar age on their respective pairs of observables. For other parameters such as [Fe/H], $Y$, and $\alpha_{\rm th}$, their effects are separable but covariant with the effect of stellar age. Finally, in the case of $f_{\rm ov,\;H\;core}$ and $f_{\rm ov,\;He\;core}$ (not shown), these observables do not cleanly separate the parameters from stellar age.

However, we wish to emphasize that Figure~\ref{fig:separation_summary}, though it is useful for illustrative purposes, does not fully encapsulate the sheer amount of information that is present in Figures~\ref{fig:vary_metallicity} through \ref{fig:vary_mdoteta}. What these panels do not capture are the subtle yet qualitatively distinct morphologies in the panchromatic CMDs and the changes to the relative number densities of stars along the CMD---in the sense of a Hess diagram---due to the model parameters. For instance, the complexities of the CMDs are boiled down to two scalar quantities, the $V-I$ difference between the MSTO and the RGB or the bluest extend of the MSTO in $B-V$, in this diagram. As such, although Figure~\ref{fig:separation_summary} appears to suggest that it is virtually impossible to infer $f_{\rm ov,\;H\;core}$ from observations, Figure~\ref{fig:vary_fovHcore} demonstrates that the detailed shape of the Henyey hook at a fixed stellar age can be used to constrain $f_{\rm ov,\;H\;core}$. This underscores the potential of a full CMD-fitting approach (e.g., MATCH, \citealt{Dolphin2002}; BASE9, \citealt{VonHippel2006}) in the era of high-precision data.

\newpage
\section{Open Clusters: Case Studies}
\label{section:case_studies}
Now that we have qualitatively explored the effects of key uncertain parameters on several observables, we evaluate the current state of the available data and assess whether they can be used to disentangle and constrain the parameters under consideration. For this exercise, we investigate three open clusters, NGC6819, M67, and NGC6791. We chose these systems for several reasons. First, all three clusters have been studied extensively and therefore represent the some of the best-case scenarios. All three clusters have been observed by the {\it Kepler} mission either as part of the original campaign (NGC6819 and NGC6791) or the repurposed {\it K2} mission (M67). They have all been observed in several photometric bands, targeted by the {\it APOGEE} spectroscopic survey \citep{Holtzman2015}, and they are known to harbor one or more DEBs. Second, non-solar-scaled abundances and multiple stellar populations are less of a concern in open clusters compared to globular clusters \citep{Bedin2004, Piotto2007}.\footnote{There is evidence that NGC6791 may be moderately $\alpha$-enhanced \citep{Linden2017}, but see also \cite{Carretta2007, Boesgaard2015, Ting2018}.} While one of the major distinctions between globular clusters and open clusters is believed to be the presence or the absence of multiple stellar populations, this simple dichotomy is becoming increasingly challenged \citep[see ][]{Gratton2012, Geisler2012, Bragaglia2014, Bastian2017}. Nevertheless, it is possible to model globular clusters and/or multiple populations in this context as well (see \citealt{Dotter2015} where the authors created tailored stellar interior and atmosphere models for NGC6752 taking into account the individual abundances of two stellar populations). The nearby globular cluster M4 may be an interesting candidate for analysis when the $\alpha$-enhanced MIST models become available, although its strong total and differential reddening \citep[e.g.,][]{Hendricks2012} may pose a challenge; it has at least three known double-lined DEBs \citep{Kaluzny2013}, {\it K2} asteroseismology \citep{Miglio2016}, and {\it APOGEE} spectra \citep{Zasowski2017}. Third, these three clusters form a sequence in age and thus allow for model comparison in different stellar mass regimes. We note that NGC6791 is noticeably more metal-rich compared to the other two clusters. 

\subsection{Cluster Membership and Distances with Gaia DR2 Data}
\label{section:gaia_mp}
We make use of the recent {\it Gaia} Data Release 2 (DR2; \citealt{Gaia2018}) to identify likely cluster members and obtain clean CMDs. First, we construct a parent catalog using the Large Survey Database (LSD; \citealt{Juric2012}) framework to combine the DR2 catalog with data from other surveys, such as {\it Pan-STARRS} \citep{Chambers2016, Flewelling2016}. We use a 3\arcsec{} threshold to cross-match the objects, selecting the nearest candidate in the event of duplicate matches. Next, we utilize the sky positions and estimates of the cluster size from the \cite{Kharchenko2013} Milky Way Star Clusters catalog to perform a cone-search centered on each cluster. In particular, we choose a search radius of twice the total apparent radius ($r_2$) to include as many of the potential cluster members as possible. 

Next, for each cluster, we identify the likely cluster members by running the HDBSCAN clustering algorithm \citep{Campello2013} on proper motion $(\mu_{\alpha}, \mu_{\delta})$ and parallax $\varpi$. Following \cite{Lindegren2018} and \cite{Babusiaux2018}, we first remove objects with failed astrometric solutions $(\texttt{\footnotesize astrometric\_chi2\_al}/({\texttt{\footnotesize astrometric\_n\_good\_obs\_al}-5})<1.44{\rm MAX}(1, \exp(-0.4(\texttt{\footnotesize phot\_g\_mean\_mag}-19.5))$). HDBSCAN identifies clusters based on the density of points and, importantly, does not force all data points to belong to a detected cluster. Unlike the DBSCAN algorithm from which it is based, HDBSCAN is more flexible in that it allows the density of the clusters to vary. Its other advantages are that there is only one important and relatively intuitive free parameter, the minimum cluster size (30), and that the algorithm returns a membership probability for every data point. We select likely cluster members with HDBSCAN membership probabilities greater than 50\% for NGC6819 and M67. For NGC6791, we relax this threshold to 30\% to retain fainter stars with lower quality astrometric data and thus increase sampling below the MSTO. 

As discussed in \cite{Luri2018}, a full Bayesian inference is the preferred method for obtaining parallax-based distances given the non-linearity between the desired (distance) and measured (parallax) quantities and the constraint that the former be necessarily positive while the latter is allowed to be zero or negative. However, as discussed in \cite{Babusiaux2018}, estimating the distance through a simple inversion is acceptable as long as the relative precision in parallax is lower than $\sim 20\%$. Following \cite{Babusiaux2018}, we adopt an even more strict 10\% precision criterion to obtain a subsample from the likely members as identified by HDBSCAN. Although this biases against fainter members in the cluster, this is not a concern for this work because completeness is not a priority. 

Before we invert the parallax measurements, we must first consider the effect of zero point offsets in {\it Gaia} DR2, which unfortunately are not well-characterized below the level of $\sim 0.05$~mas at this time \citep{Arenou2018, Lindegren2018}. Independent comparisons with quasars \citep{Lindegren2018}, external catalogs and Milky Way satellites \citep{Arenou2018}, eclipsing binaries \citep{Stassun2018}, and RR Lyrae \citep{Muraveva2018} have revealed both large- and small-scale ($<1^{\circ}$) spatial variations in the parallax zero points, with an average global offset of $\sim-0.03 \textrm{ -- } -0.08$~mas ({\it Gaia} parallaxes are systematically too small). We cannot reliably correct for this by adding a constant zero point offset given that the small-scale variations are comparable in size to the average global offset. As we describe in detail in the following sections, this can lead to differences in the absolute distance modulus by a value as large as $\sim0.4$~mag in the case of NGC6819. For this work, we first compute the fiducial cluster distance using the median parallax from the high-precision subsample, apply the resulting distance modulus to the model CMDs, and assess the quality of the fits to all of the available cluster data. If the resulting fit is entirely inconsistent with the observations (i.e., impossible to reconcile by changing the age or the reddening), then we allow for a small increase in the parallax by an amount between 0 and $0.08$~mas. In a future work, we will present a detailed description of the cluster membership selection process in addition to a more rigorous determination of the cluster parameters using a hierarchical Bayesian analysis where we also model the parallax zero points.

\begin{figure*}
\centering
\includegraphics[width=0.85\textwidth]{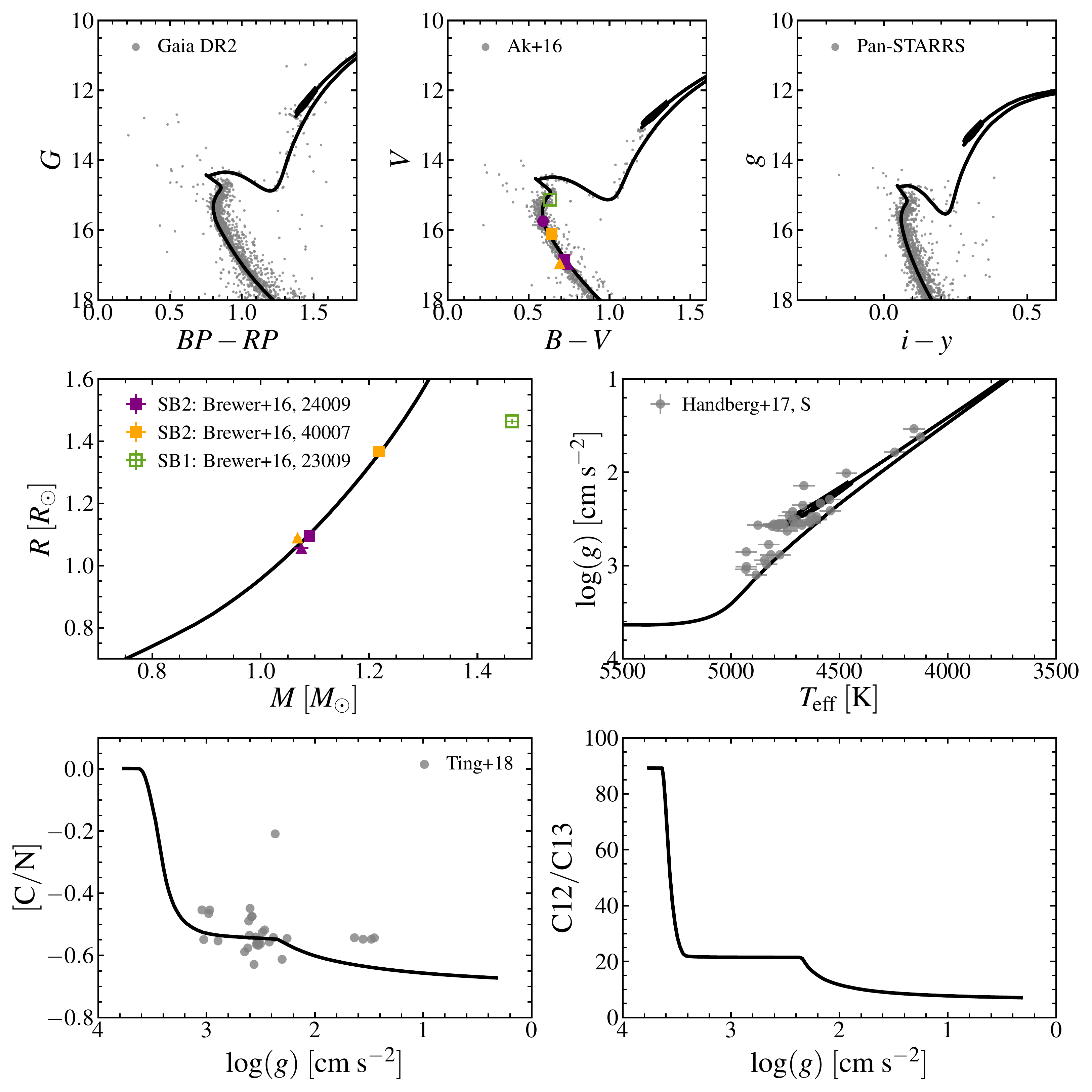}
\caption{NGC6819, a solar-metallicity, intermediate-age (2~Gyr) open cluster. The top three panels show photometry of likely cluster members selected using the {\it Gaia} DR2 data, along with the MIST models (see text for more details). The MIST CMDs adopt $m-M=11.8$, $\rm [Fe/H]=-0.01$,  $\logage=9.4$, and $A_V=0.5$. The left panel in the middle row shows stellar parameters measured from DEB \citep{Brewer2016} with the MIST mass-radius relation. The open and closed symbols indicate that the system is a single- (SB1) and double-lined (SB2) spectroscopic binary system, respectively. When available, the individual EB components are also shown in the CMDs, where the square, triangle, and circle symbols correspond to the primary, secondary, and tertiary components, respectively. The right panel shows the asteroseismic $\logg$ of single stars inferred from the scaling relations \citep{Handberg2017} with the effective temperatures from the \cite{Casagrande2014} color-temperature relations. The bottom left panel shows the comparison between predicted surface [C/N] abundances with the measured abundances, shown as gray circles. The surface abundances are obtained by reanalyzing the publicly-available {\it APOGEE} DR14 spectra \citep{Holtzman2015, SDSS2016} with {\it the Payne} \citep{Ting2018}. The measurement of $\rm ^{12}C/^{13}C$ in the right panel is currently being explored.}
\label{fig:ngc6819}
\end{figure*}

\begin{figure*}
\centering
\includegraphics[width=0.85\textwidth]{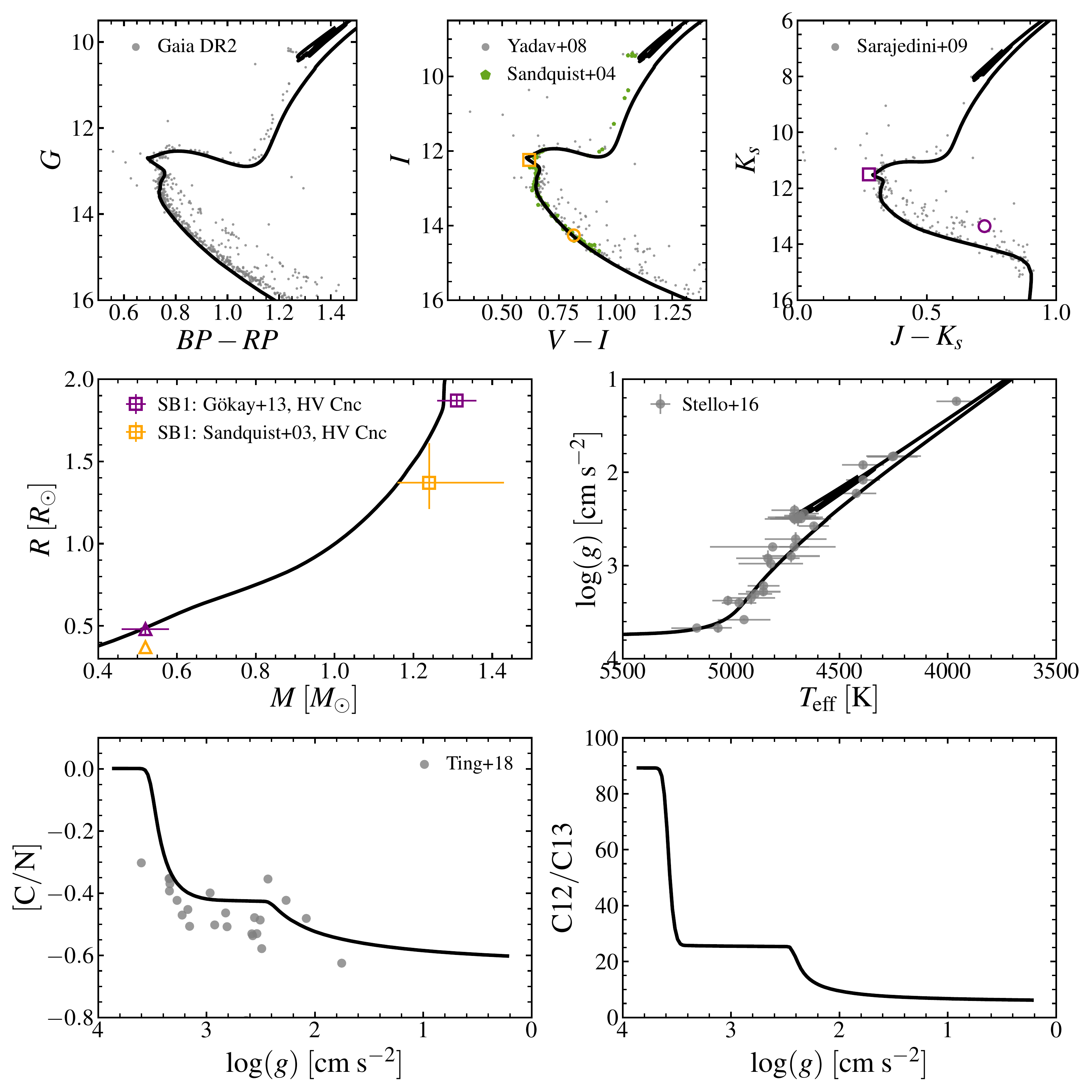}
\caption{Same as Figure~\ref{fig:ngc6819} except now for M67, a solar-metallicity and solar-age (4~Gyr) open cluster. The top left panel shows photometry of likely cluster members selected using the {\it Gaia} DR2 data while the middle and right panels show stars with $>50\%$ membership probabilities as determined by \cite{Yadav2008}. The \cite{Sandquist2004} is a sample of likely single star members selected based on their proper motions. The MIST CMDs, shown in black, adopt $m-M=9.73$, $\rm [Fe/H]=-0.01$, $\logage=9.58$, and $A_V=0.18$ (see text for more details). The left panel in the middle row shows two measurements of the stellar parameters of HV Cnc \citep{Sandquist2003, Goekay2013} with the MIST mass-radius relation. The open symbols indicate that the system is a single-lined (SB1) spectroscopic binary system. When available, the individual EB components are also shown in the CMDs, where the square, triangle, and circle symbols correspond to the primary, secondary, and tertiary components, respectively. The right panel shows the asteroseismic $\logg$ inferred from the scaling relations \citep{Stello2016} with the effective temperatures from the \cite{Casagrande2014} color-temperature relations. The surface abundances are obtained by reanalyzing the publicly-available {\it APOGEE} DR14 spectra \citep{Holtzman2015, SDSS2016} with {\it the Payne} \citep{Ting2018}. The measurement of $\rm ^{12}C/^{13}C$ in the right panel is currently being explored.}
\label{fig:m67}
\end{figure*}

\subsection{NGC6819}
NGC6819 is a solar-metallicity, intermediate-age (2~Gyr), richly populated open cluster \citep{Yang2013, AnthonyTwarog2014, LeeBrown2015}. As the youngest system in our sample, its MSTO stars are massive enough to have convective cores, giving rise to a distinctive MSTO morphology called the Henyey hook.

Figure~\ref{fig:ngc6819} shows the multi-panel plot summary of NGC6819. The panels are analogous to those presented in Figures~\ref{fig:vary_metallicity} through \ref{fig:vary_mdoteta}. The photometry comes from several sources: {\it Gaia} DR2 \citep{Evans2018}, $BVI$ \citep{Ak2016}, and {\it Pan-STARRS} \citep{Chambers2016, Flewelling2016}. We utilize the {\it Gaia} DR2 membership (see Section~\ref{section:gaia_mp}) to select the likely cluster members from the \cite{Ak2016} catalog. We apply additional cuts in the {\it Gaia} photometry using \texttt{phot\_bp\_rp\_excess\_factor} following \cite{Babusiaux2018}. The reported photometric uncertainties are $\sim 2$, 3, and 5~mmag in $G$, $V$, and $g_{\rm PS}$ near the MSTO. Note that the {\it Pan-STARRS} CMD is truncated above $g_{\rm PS}\approx15$ due to saturation, but nevertheless we include it for its well-sampled lower MS and availability of the NIR photometry. The black isochrone corresponds to an example ``fit'' to the data using the MIST isochrone. The metallicity is assumed to be the median [Fe/H] value for the sample of red giants, measured from the publicly-available {\it APOGEE} DR14 spectra \citep{Holtzman2015, SDSS2016} with {\it the Payne} (\citealt{Ting2018}; see below for more details) and the fiducial value of the distance modulus is estimated from the {\it Gaia} DR2 parallax measurements as described in Section~\ref{section:gaia_mp}. Finally, we start with literature values of reddening and age and choose the ``best-fit'' values by eye for purely illustrative purposes. However, we found that $m-M=12.22$, the distance modulus inferred from the {\it Gaia} DR2 parallax---and the subsequent adjustments to the age and reddening in order to fit the MS in the CMDs---is strongly ruled out by both the RC magnitude and the rest of the observations. This is consistent with the conclusion from the literature that the {\it Gaia} DR2 parallaxes are systematically too small (see the discussion in Section~\ref{section:gaia_mp}). As such, we are afforded some flexibility to apply a zero point offset to the fiducial cluster parallax, and we find that 0.075 mas is a suitable value, which, though large, is still within the range of estimates from the literature. In summary, the MIST models assume $m-M=11.8$, $\rm [Fe/H]=-0.01$,  $\logage=9.4$, and $A_V=0.5$.

The mass and radius measurements of the two EB systems, WOCS~24009 (Auner~665; KIC~5023948) and WOCS~40007 (KIC~5113053), are derived from a combination of {\it Kepler} and ground-based photometry and spectroscopy \citep{Brewer2016}. In fact, each one belongs to its own triple system: WOCS~24009 is a triple-lined system where the brightest, non-eclipsing component is orbiting a short-period binary system and WOCS~40007 is a double-lined system. There is a third EB system WOCS~23009 \citep{Hole2009, Sandquist2013}, but it is a single-lined EB and thus the inferred parameters are less certain. Nevertheless, we include the parameters of the primary in our comparison, but in open symbols to reflect its lower fidelity. When available, the individual EB components are also shown in the CMDs, where the square, triangle, and circle symbols correspond to the primary, secondary, and tertiary components, respectively.

Moving on to the right panel, we show the asteroseismic $\logg$ and $\teff$ for single stars from \cite{Handberg2017}. The authors used a variation of the classic scaling relations that are recast to include bolometric luminosities. The bolometric corrections and color-temperature relations that are required to estimate $L$ and $\teff$ come from \cite{Casagrande2014}. They utilize $V$ \citep{Milliman2014} and $Ks$ \citep{Cutri2003} photometry and assume a nominal reddening value of $E(B-V)=0.15$ and [Fe/H]=0.02$\pm0.10$. We adopt a $\teff$ uncertainty of 50~K following the authors' estimates. Overall, the asteroseismic $\logg$ and $\teff$ are in good agreement with the MIST model predictions, in particular the RC magnitude and the RGB $\teff$. 

Finally, the last row shows the comparison between predicted surface [C/N] abundances with measured the observed abundances, shown as gray circles. The surface abundances are provided by {\it the Payne} \citep{Ting2018} re-analysis of the {\it APOGEE} DR14 spectra \citep{Holtzman2015, SDSS2016}. Here we only briefly describe {\it the Payne} since the details of the methodology are presented in \cite{Ting2018}. {\it The Payne} utilizes the idea of generative models; it fits the variations in normalized flux with respect to stellar labels (stellar parameters and elemental abundances) with a flexible functional form approximated with neural networks. The neural networks are trained on the ATLAS12/SYNTHE model spectra \citep{Kurucz1970, Kurucz1981, Kurucz1993} and the observed spectra are fit via full-spectral fitting. The formal uncertainties for {\it the Payne} are very small ($< 0.01$~dex), but the true uncertainties are usually dominated by model systematics. The abundance spreads measured in open clusters, which are presumed to be chemically homogeneous, imply a precision of $\approx0.03$~dex \citep{Ting2018}. {\it The Payne} [C/N] abundances are in good agreement with the MIST prediction. The full-spectral fitting approach in principle allows for the measurement of $\rm ^{12}C/^{13}C$, which is currently being explored. 

\begin{figure*}
\centering
\includegraphics[width=0.85\textwidth]{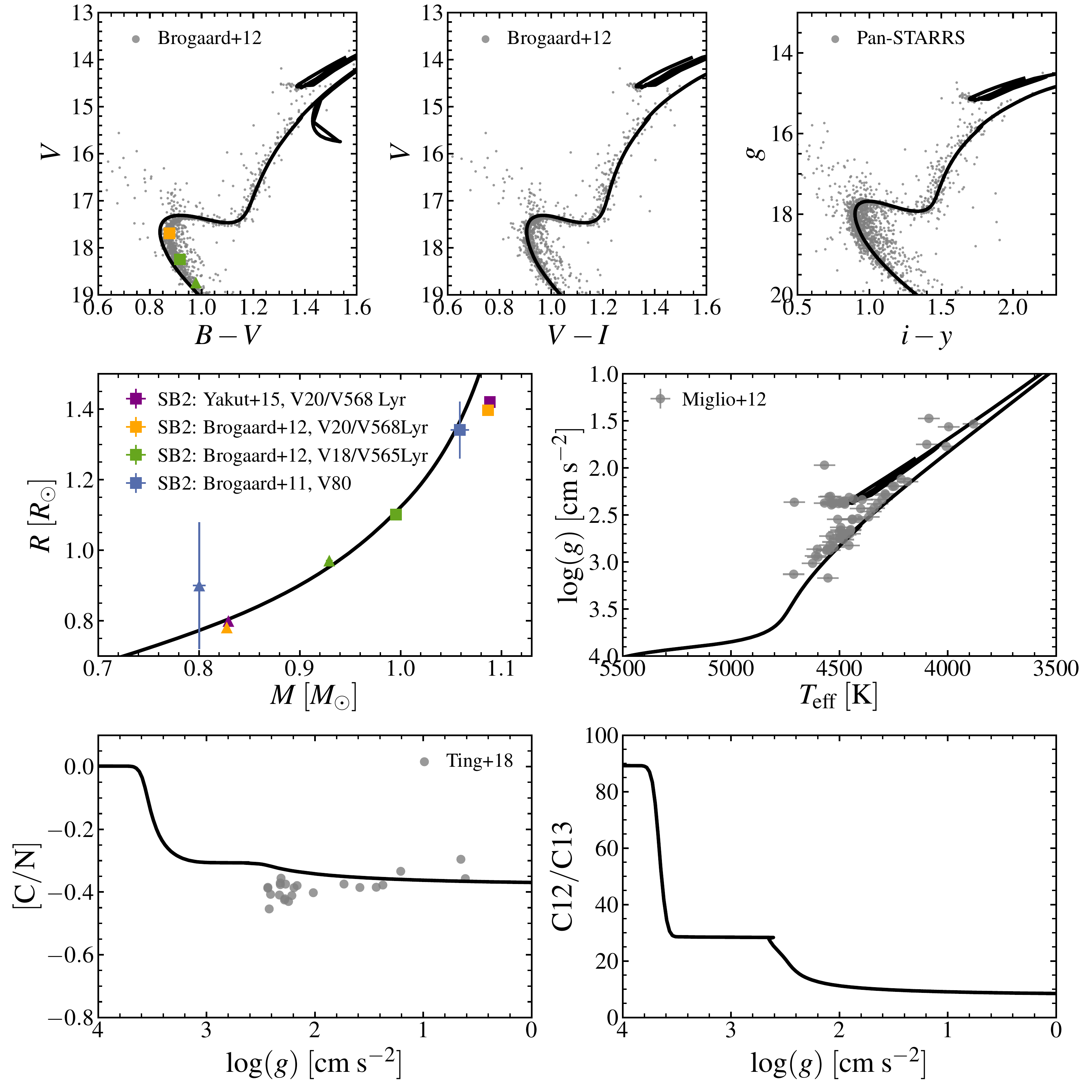}
\caption{Same as Figure~\ref{fig:ngc6819} except now for NGC6791, a metal-rich ($\rm [Fe/H] \approx +0.3$) and old (8~Gyr) open cluster. The top left panel shows photometry of likely cluster members selected using the {\it Gaia} DR2 data. The MIST CMDs assume $m-M=13.05$, $\rm [Fe/H]=+0.28$, $\logage=9.95$, and $A_V=0.434$, and show interesting tension with the data (see the text for more details). The left panel in the middle row compares the mass and radius measurements of three DEB systems \citep{Brogaard2011, Brogaard2012, Yakut2015} with the MIST mass-radius relation. The closed symbols indicate that the system is a double-lined (SB2) spectroscopic binary system. When available, the individual EB components are also shown in the CMDs, where the square and triangle symbols correspond to the primary and secondary, respectively. The right panel shows the asteroseismic $\logg$ inferred from the scaling relations \citep{Miglio2012} with the effective temperatures from the \cite{Ramirez2005} color-temperature relations. The surface abundances are obtained by reanalyzing the publicly-available {\it APOGEE} DR14 spectra \citep{Holtzman2015, SDSS2016} with {\it the Payne} \citep{Ting2018}. The measurement of $\rm ^{12}C/^{13}C$ in the right panel is currently being explored.}
\label{fig:ngc6791}
\end{figure*}

\subsection{M67}
M67 is a nearby solar-metallicity, intermediate-age (4~Gyr) open cluster \citep{Taylor2007, Sarajedini2009, Onehag2014}. One of the reasons it is so well studied is that its MSTO mass is very close to the transition mass above and below which stars burn hydrogen convectively and radiatively in their cores. Its Henyey hook is frequently used to calibrate the efficiency of convective overshoot mixing in low mass stars \cite[e.g.,][]{VandenBerg2006, Magic2010, Bressan2012, Choi2016}.

Figure~\ref{fig:m67} shows the multi-panel plot summary of M67. The optical and near-infrared photometry comes from {\it Gaia} DR2 \citep{Evans2018}, \cite{Yadav2008}, and \cite{Sarajedini2009}. For the latter two catalogs, we only show a subset of the stars with membership probabilities greater than $50\%$, as evaluated by \cite{Yadav2008} using their own proper motion data. An additional proper motion selected sample of likely single-star members from \cite{Sandquist2004} is overplotted for comparison. The reported photometric uncertainties are $\sim 4$, 3, and 5~mmag in $G$, $I$, and $Ks$ near the MSTO. Similar to the approach used to model NGC6819, we adopt the [Fe/H] value from {\it the Payne} and estimate the distance modulus from the {\it Gaia} DR2 parallax. We do not apply a parallax zero point offset, though an increase of 0.029~mas (the global zero point as estimated in \citealt{Lindegren2018}), resulting in a decrease in the distance modulus of $\sim 0.05$~mag is also permissible with a small decrease in age. In summary, the MIST models assume $m-M=9.73$, $\rm [Fe/H]=-0.01$, $\logage=9.58$, and $A_V=0.18$. In $BP-RP$, the model diverges from the data near $G\lesssim15$ for reasons that are unknown at this time. Interestingly, $B-V$ and $B-I$ (not shown) show excellent agreement on the lower MS down to $G\approx16.5$, below which the model diverges blueward of the data due to the well-known M-dwarf radius inflation problem \citep[e.g.,][]{Kraus2011, Feiden2013, Torres2013}. Moreover, there is moderate tension between the predicted and observed RGB colors at the $\sim 0.05$~mag level in $BP-RP$ and $V-I$, which may point to interesting model deficiencies (e.g., $\amlt$ variation; \citealt{Bonaca2012, Tayar2017} but see also \citealt{Salaris2018}; \citealt{Choi2018}) and/or poorly-``fit'' cluster parameters, but we do not conclusively attribute the discrepancy to any one source at this time.

The left panel in the middle row compares the MIST mass-radius relation with two independent mass and radius determinations of HV~Cnc. HV~Cnc was initially reported to be a single-lined binary \citep{Mathieu1990}, but detections of a weak secondary and a possible tertiary component were reported in subsequent works \citep{Melo2001, Sandquist2003}. \cite{Sandquist2003} analyzed $VI$ photometry and radial velocity data and found a third non-binary component in the spectra, though its association with the HV~Cnc system is still uncertain. They deconvolved the photometry of the three stars to yield the parameters of the two binary components, shown in yellow. The primary is hotter than the majority of the cluster MSTO stars, which suggests that it is either a blue straggler or undergoing the overall contraction phase along the Henyey hook. \cite{Goekay2013} provided an updated set of parameters by adding in the $JHKs$ photometry, confirming the spectroscopic detection of a third component. They combined the radial velocity solution of the primary with the mass ratio inferred from the light curves in order to obtain the full solution of the binary system, shown in purple points. Again, we use open symbols to indicate that HV Cnc is a single-lined binary system. When available, the individual EB components are also shown in the CMDs, where the square, triangle, and circle symbols correspond to the primary, secondary, and tertiary components, respectively.

The right panel shows the asteroseismic $\logg$ from the analysis of the {\it K2} photometry \citep{Stello2016}. The authors computed $\teff$ using the optical and 2MASS photometry with the color-temperature relations from \cite{Casagrande2010}, assuming $\rm [Fe/H] = 0$ and $E(B-V)=0.03$. Their $\teff$ uncertainties were estimated by the scatter in the $\teff$ inferred from different combinations of the photometric systems, plus an additional 20~K to account for the $\teff$ zero point uncertainty. Overall, the $\teff$ and $\logg$ are in excellent agreement with the model predictions including the RC magnitude, albeit the $\teff$ uncertainties are quite large.

Finally, we plot the [C/N] and $\rm ^{12}C/^{13}C$ surface abundance evolution on the RGB. {\it The Payne} [C/N] abundances are shown as gray circles for comparison. The spectroscopic [C/N] abundances fall $\approx 0.1$~dex below the MIST prediction, suggesting a weak preference for a younger age and/or higher metallicity (see Figure~\ref{fig:vary_metallicity}). Measurement of $\rm ^{12}C/^{13}C$ from the {\it APOGEE} spectra is currently being explored.

\begin{figure*}
\centering
\includegraphics[width=0.8\textwidth]{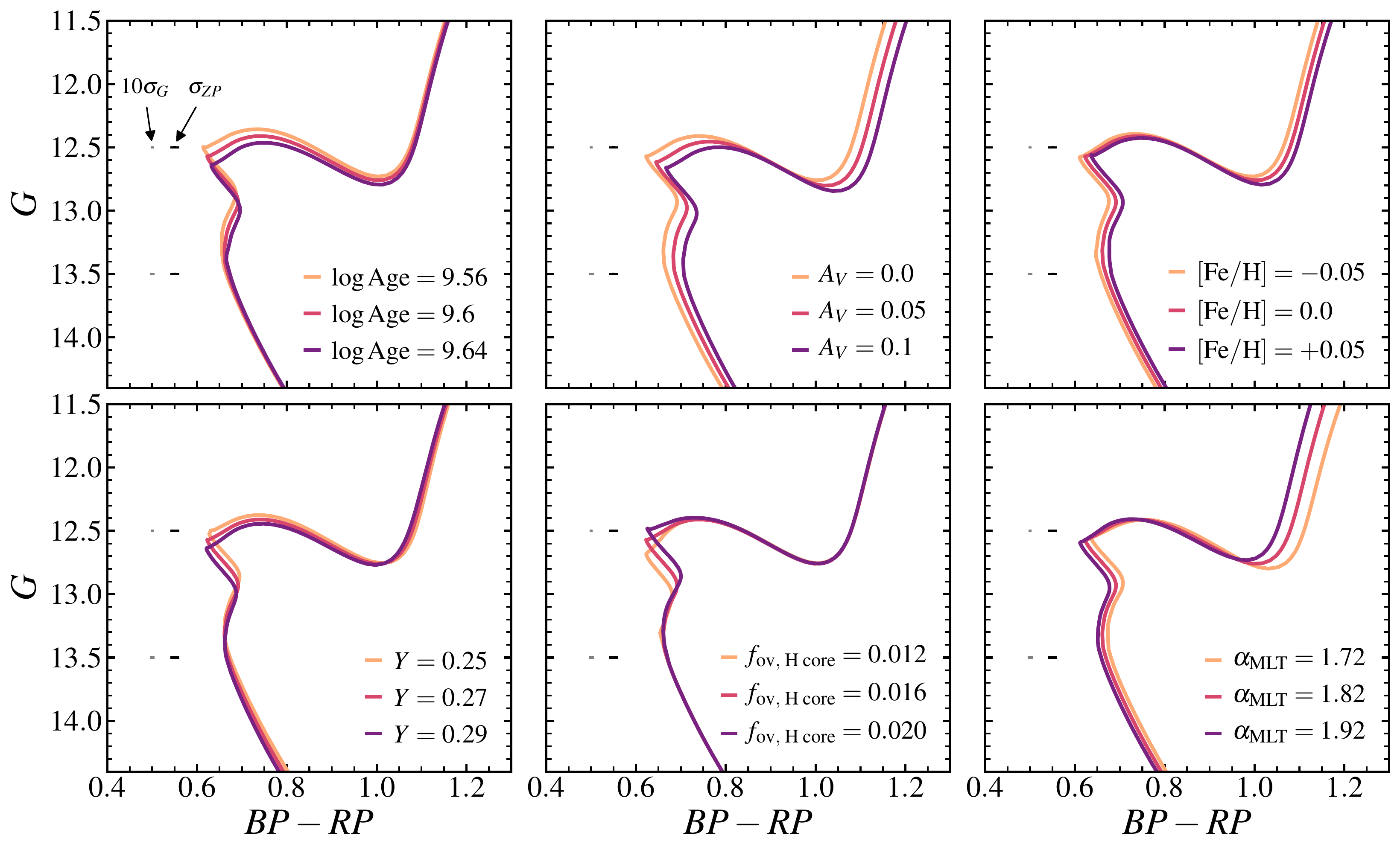}
\caption{MIST isochrones that illustrate the effects of uncertain parameters on various parts of the {\it Gaia} CMD. A distance modulus of $m-M = 9.7$ is applied to all models and extinction is not included unless noted otherwise. The $\log \rm Age$, [Fe/H], initial helium abundance, convective overshoot efficiency in the hydrogen-burning core, and mixing length $\alpha$ are held fixed to 9.6, 0.0, 0.2703, 0.016, and 1.82 unless noted otherwise. While these parameters indeed have only a subtle influence on the CMD morphology, they change the CMD in unique ways and thus should be separable with high quality models, data, and fitting tools. What these panels do not explicitly show is the effect of these parameters on the lifetimes. The representative {\it Gaia} end-of-mission median-straylight photometric standard errors assuming 70 visits per field are shown in gray. For display purposes, we multiply the uncertainties in each band by a factor of 10. We also show the absolute photometric accuracy due to zero point uncertainty ($\approx 0.014$~mag) in black. Top left: varying the stellar age. Top middle: varying the amount of reddening assuming the $R_V=3.1$ reddening law from \cite{Cardelli1989}. Top right: varying the metallicity. Bottom left: varying the initial helium abundance. Bottom middle: varying the convective overshoot mixing in the hydrogen core. Bottom right: varying the mixing length $\amlt$.}
\label{fig:gaia_error}
\end{figure*}

\subsection{NGC6791}
NGC6791 is an exceptionally old (8~Gyr) and metal-rich ([Fe/H]$\approx$0.3--0.5) open cluster \citep[e.g.,][]{Stetson2003, Gratton2004, Carney2005, King2005, Origlia2006, Linden2017}. It is also well-known for its puzzling double white dwarf cooling sequence, both of which imply cluster ages that are nominally inconsistent with the MSTO age (\citealt{Bedin2005, Bedin2008a} but see also \citealt{GarciaBerro2010}). Several explanations have been put forth, including the presence of a secondary population of massive helium WDs \citep{Hansen2005, Kalirai2007} and WD binaries \citep{Bedin2008b}. 

Figure~\ref{fig:ngc6791} shows the multi-panel plot summary of NGC6791. The left and middle panels show $BVI$ photometry from \cite{Brogaard2012}, which includes differential reddening corrections to the \cite{Stetson2003} photometry, while the right panel shows {\it Pan-STARRS} photometry \citep{Chambers2016, Flewelling2016}. All three panels show likely members selected by the {\it Gaia} DR2 proper motion and parallax data. We apply additional quality cuts in the \cite{Brogaard2012} data, restricting the sample to those with $\sigma_{V} < 10$~mmag. We do not show the {\it Gaia} DR2 photometry because it is too shallow and shows a much larger scatter. The reported photometric uncertainties are $\lesssim 1 \textrm{--}10$ and 6~mmag in $V$ and $g_{\rm PS}$ near the MSTO, respectively.\footnote{The photometric errors in $V$ near $V\approx18$ as reported by \cite{Brogaard2012} show a bimodal distribution with peaks at $\sigma_V \approx 0.3$ and 5~mmag, respectively, after we limit the sample to $\sigma_V<10$~mmag. The MS is not much different when we restrict the sample to $\sigma_V<1$~mmag.} As before, the metallicity is fixed to the {\it the Payne} measurement. The distance modulus estimated from the {\it Gaia} DR2 parallaxes is uncertain, which is unsurprising given the large distance ($\sim 4$~kpc). If we compute the distance modulus by adopting the $10\%$ precision cut (21 stars total) and calculating the median parallax as described in Section~\ref{section:gaia_mp}, we obtain $m-M=13.29$. However, if we relax the precision cut and/or calculate the mean or the weighted median rather than the median, the resulting distance modulus can range from $\sim13.03$ to $\sim 13.45$. This is expected to improve with more data over the next several years, but for now, we start with $m-M=13.29$ and allow flexibility in the final adopted value. In summary, the MIST models, shown in black, assume $m-M=13.05$ (requiring a zero point offset of $0.025$~mas), $\rm [Fe/H]=+0.28$, $\logage=9.95$, and $A_V=0.434$. Interestingly, these models cannot simultaneously fit the MSTO in all three CMDs and they also show moderate discrepancy in the RGB color. A simple increase in the reddening value improves the agreement in $B-V$, but not without introducing tension in $V-I$ and $i-y$. An increase in metallicity to $+0.4$~dex results in an overall improved fit at the MSTO in the three panels, but at the cost of increasing tension on the RGB, not to mention introducing inconsistency with {\it the Payne} metallicity.

In the middle row, we compare the mass-radius measurements of three systems, V18, V20, and V80, with the MIST model predictions. \cite{Brogaard2012} updated the analysis for V18 and V20 from \cite{Brogaard2011} using a new photometric reduction procedure and an improved analysis of the V20 secondary: the contribution to the light curve from the third component in V20 was accounted for using four ``twin stars'' that were identified in a much higher-resolution {\it HST}/ACS image. According to \cite{Brogaard2011}, photometric and radius measurements of V80 are very uncertain due to magnetic activity possibly induced by its close-in orbit. Nevertheless, all three systems are double-lined binaries: there are mass measurements for all three systems and radius measurements for V18 and V20 to within $1\%$. Additionally, there is an updated measurement for V20 from \cite{Yakut2015} where the authors utilized very precise {\it Kepler} light curves to obtain more accurate estimates of the stellar parameters. The most massive EB shows tension with the isochrone, requiring either a younger age at fixed [Fe/H] or a higher [Fe/H] at fixed age to resolve the discrepancy. When available, the individual EB components are also shown in the CMDs, where the square and triangle symbols correspond to the primary and secondary, respectively. 

In the next panel, we plot the asteroseismic $\logg$ measurements from \cite{Miglio2012}. The authors adopted effective temperatures calculated from the \cite{Ramirez2005} $V-K$ color-temperature relation assuming $\rm [Fe/H] = +0.3$ and $E(B-V)=0.16\pm0.02$. Following \cite{Hekker2011}, they assume an uncertainty of 50~K, though they caution that systematic uncertainties due to color-temperature calibrations and reddening could result in a number closer to $\sim 110$~K. We adopt $50$~K for computing uncertainties in $\logg$. There is a mild offset of $\lesssim 50$~K between the predicted and ``observed'' temperatures, though note that the latter was estimate using a reddening value that is larger than what is adopted for the CMDs.

In the last row, we plot the [C/N] and $\rm ^{12}C/^{13}C$ surface abundance evolution on the RGB. For comparison, {\it the Payne} [C/N] abundances are shown as gray circles. Finally, we plot the [C/N] and $\rm ^{12}C/^{13}C$ surface abundance evolution on the RGB. {\it The Payne} [C/N] abundances are shown as gray circles for comparison, which are generally in good agreement with the MIST prediction. Measurement of $\rm ^{12}C/^{13}C$ from the {\it APOGEE} spectra is currently being explored.

\newpage
\section{What We Can Expect From Gaia In The Future}
\label{section:gaia}
The {\it Gaia} mission was designed to obtain $\mu$as astrometry and proper motions for a billion Milky Way stars along with high-precision photometry consisting of both broadband $G$ and blue/red ($B/R$) spectrophotometry \citep{Jordi2010}. While this work already benefited greatly from {\it Gaia} DR2, especially in cluster membership, there should be notable improvement with future data releases in e.g., zero points and uncertainties in the astrometric solutions. In this section, we illustrate what we might expect from end-of-mission {\it Gaia} data using M67 as a fiducial case.

In Figure~\ref{fig:gaia_error}, we show example CMDs representative of M67, where each of the six panels shows a series of MIST models illustrating the effects of uncertain parameters. The gray error bars represent the end-of-mission (assuming 70 visits to each field) photometric standard errors estimated according to a performance model made available by the {\it Gaia} mission.\footnote{\url{https://www.cosmos.esa.int/web/gaia/science-performance}} For display purposes, we inflate the errors in each band by a factor of 10. We emphasize that these errors are representative of {\it relative} photometric precision only, because the {\it absolute} photometric accuracy is still dominated by the photometric zero point uncertainty. The photometric zero point measurement is tied to the $1\%$ calibration of Vega's spectra (see \citealt{Carrasco2016} and also the discussion in Section~\ref{section:observations_phot}), which ultimately yields $\approx0.014$~mag absolute photometric uncertainty in color, shown as black error bars ($0.01$~mag added in quadrature). However, these uncertain model parameters change the CMD morphology in qualitatively distinct ways, and thus should be separable with high quality models, data, and fitting tools even in the presence of zero point uncertainties. What these panels do not explicitly show is the effect of these parameters on the lifetimes---this information is encoded in the density distribution or number ratios of stars in various parts of the CMD. The subtle differences illustrated in Figure~\ref{fig:gaia_error} are difficult to distinguish using traditional techniques, e.g., fitting empirical MS ridgelines, but we will soon be able to leverage the exquisite {\it Gaia} photometry, proper motions, and parallax, in combination with a diverse dataset including spectroscopy and asteroseismology. These data sets will place very stringent constraints on the models in Figures~\ref{fig:ngc6819}, \ref{fig:m67}, and \ref{fig:ngc6791}, which show isochrones that were ``fit'' by eye for illustrative purposes. A future direction in this area includes a quantitative and objective determination of the best likelihood parameters (e.g., MATCH, \citealt{Dolphin2002}; BASE9, \citealt{VonHippel2006}; MINESweeper, Cargile et al., in prep.).

\section{Summary}
\label{section:summary}
In this work, we provided an overview of the currently available and future data sets that can be leveraged simultaneously to both improve our constraints on the uncertain stellar model parameters and to infer the properties of open clusters. We first explored the effects of key parameters---age, metallicity, helium content, mixing length parameter, convective boundary mixing efficiency in hydrogen and helium cores, thermohaline mixing efficiency, and mass loss efficiency---on the various observational diagnostics. Next, we identified pairs of observables that are sensitive to each parameter of interest and stellar age, taking into account the observational feasibility. The key plot that summarizes the results is shown in Figure~\ref{fig:separation_summary}.

There are several important caveats. At this level of scrutiny, photometric/parallax zero points and differential reddening (see the discussion in Sections~\ref{section:other_parameters_to_consider} and \ref{section:gaia_mp}) may well dominate the observational uncertainties. However, the zerpoints induce an overall shift in the CMD while the key parameters considered in this work shape the CMD morphologies in qualitatively distinct ways, and thus the two types should be separable. On the theoretical modeling side, a proper treatment of the detailed abundance patterns (for example, see \citealt{Dotter2015} where the authors analyzed NGC6752 using self-consistent stellar interior and atmosphere models computed according to the detailed spectroscopic abundances), the effects of atomic diffusion on the surface abundances \citep{Dotter2017}, and the surface boundary conditions (e.g., \citealt{Salaris1996, Chabrier1997, VandenBerg2008, Choi2018}) will likely be important.

We also evaluated the current status of the various observational data sets using three well-studied open clusters---NGC6819, M67, and NGC6791---as case studies. Although we find no obvious discrepancies between the existing data and the MIST models for NGC6819 (Figure~\ref{fig:ngc6819}), M67 shows a mild tension in the RGB colors (Figure~\ref{fig:m67}). NGC6791 appears to prefer a slightly higher metallicity than what is inferred from {\it the Payne} analysis of the {\it APOGEE} spectra, though this would lead to a tension in the RGB colors (Figure~\ref{fig:ngc6791}). More precise observations (e.g., parallax-based distance for NGC6791) and robust fitting will help to conclusively identify and quantify the discrepancies. {\it Gaia} parallax measurements, with careful modeling of the zero point offsets, should remove distance as a source of uncertainty, and the accompanying {\it Gaia} photometry ($B$, $R$, and $G$; see Figure~\ref{fig:gaia_error}) and proper motion memberships will immensely improve the quality of the CMDs, as already demonstrated with the DR2 data. CMDs contain a tremendous amount of information, and thus the combination of exquisite photometry, flexible and robust stellar models, and objective fitting tools will allow us to measure stellar ages and disentangle the effects of key stellar model parameters in the near future.

\section*{Appendix}
\begin{figure*}
\centering
\includegraphics[width=0.8\textwidth]{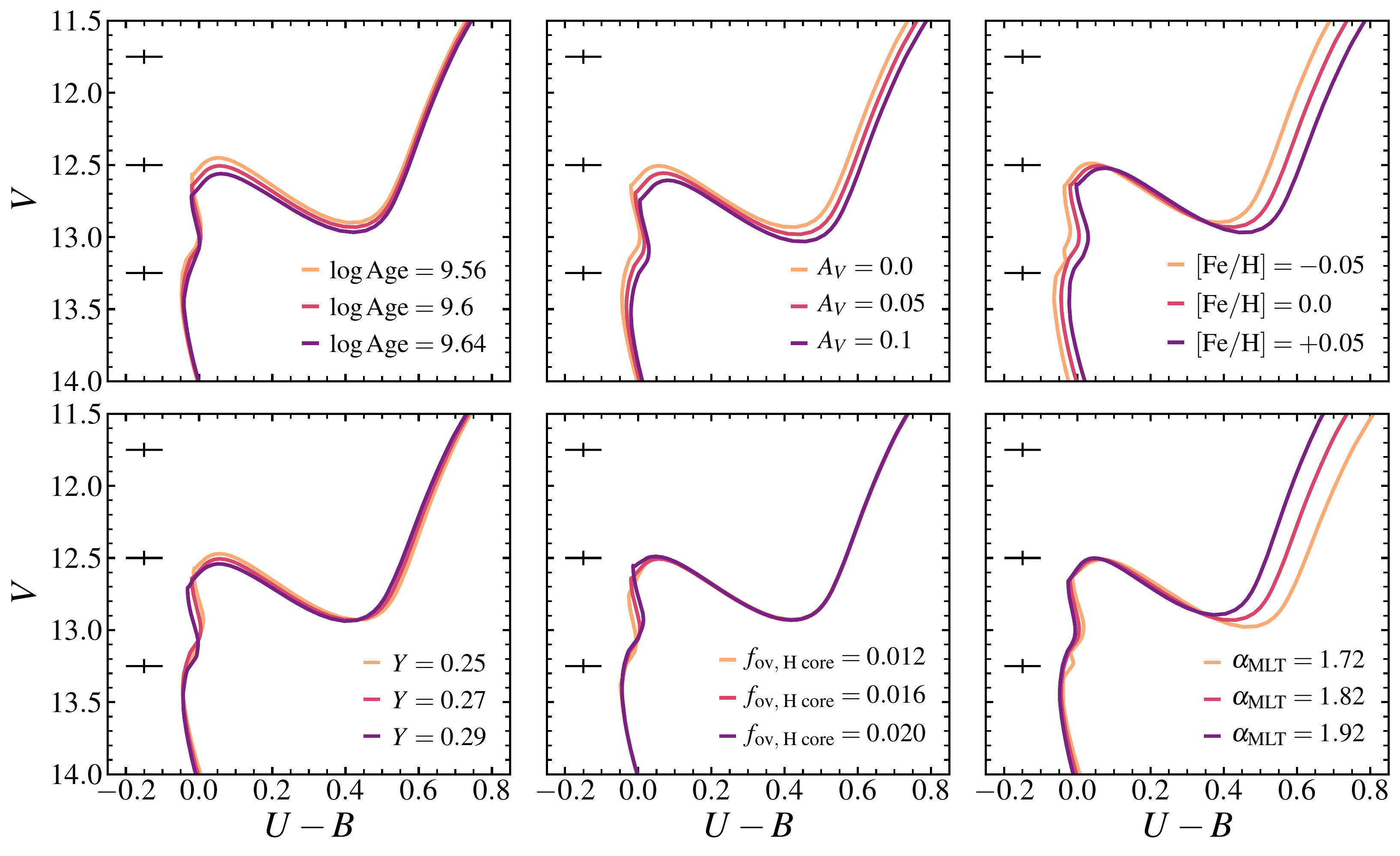}
\caption{Same as Figure~\ref{fig:gaia_error} except now showing $V$ vs. $U-B$. We also show error bars, 0.05~mag in both color and magnitude, to help guide the eye.}
\label{fig:ubv}
\end{figure*}

Here we present a series of figures illustrating the effects of uncertain parameters---$\log \rm Age$, extinction, [Fe/H], initial helium abundance, convective overshoot efficiency in the hydrogen-burning core, and mixing length $\alpha$---on various parts of the CMD. These figures are analogous to Figure~\ref{fig:gaia_error} and show qualitatively the same behavior, but they demonstrate that the sizes of the effects can vary significantly depending on the combination of filters. We also include error bars, 0.05~mag in both color and magnitude, to guide the eye in each panel and assist with the direct comparison of CMDs plotted on different axis scales. We caution the reader against over-interpreting the minor blemishes in these figures, e.g., the $\amlt=1.72$ curve in the bottom right panel of Figure~\ref{fig:ubv}.

\begin{figure*}
\centering
\includegraphics[width=0.8\textwidth]{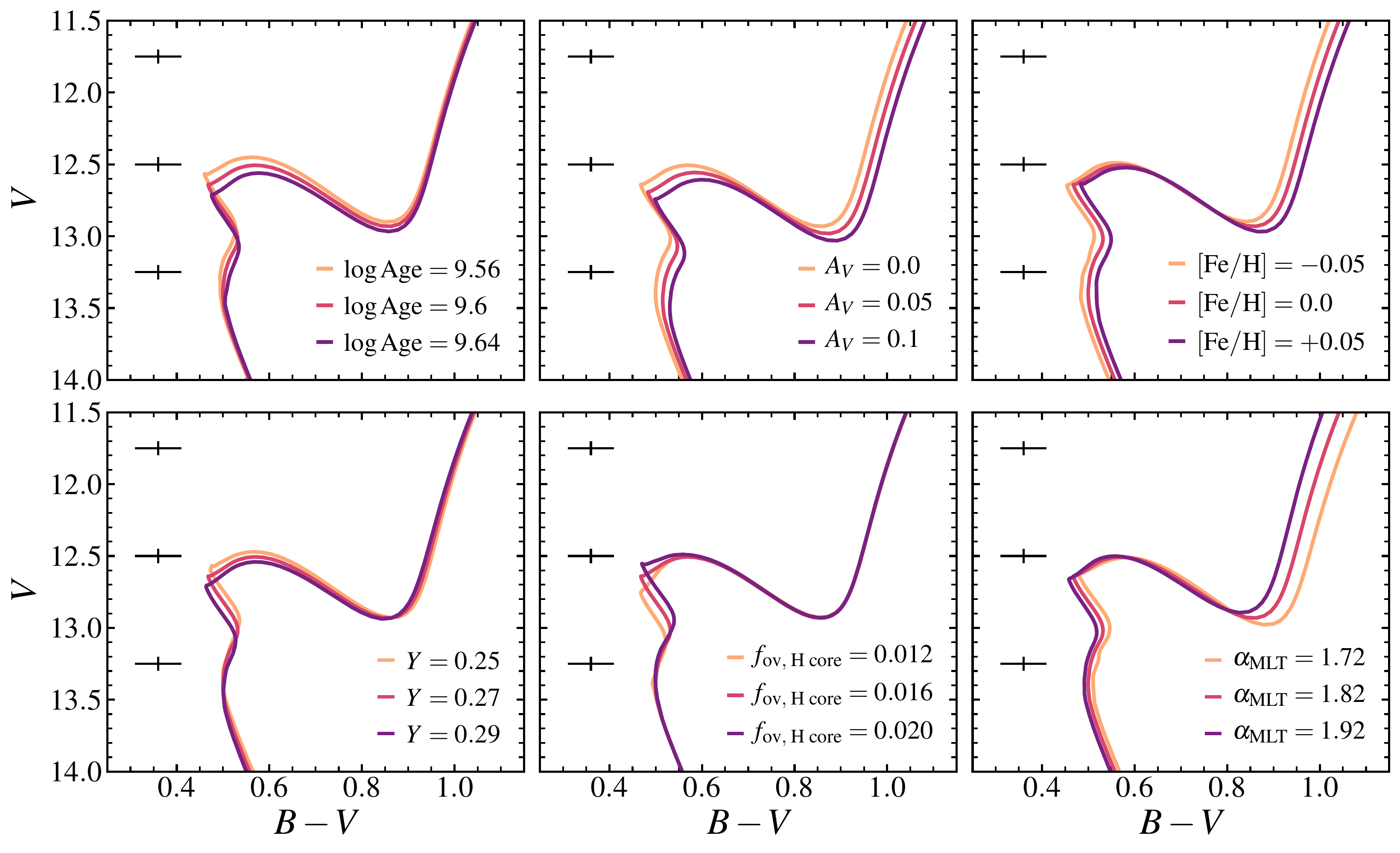}
\caption{Same as Figure~\ref{fig:ubv} except now showing $V$ vs. $B-V$.}
\label{fig:bvv}
\end{figure*}

\begin{figure*}
\centering
\includegraphics[width=0.8\textwidth]{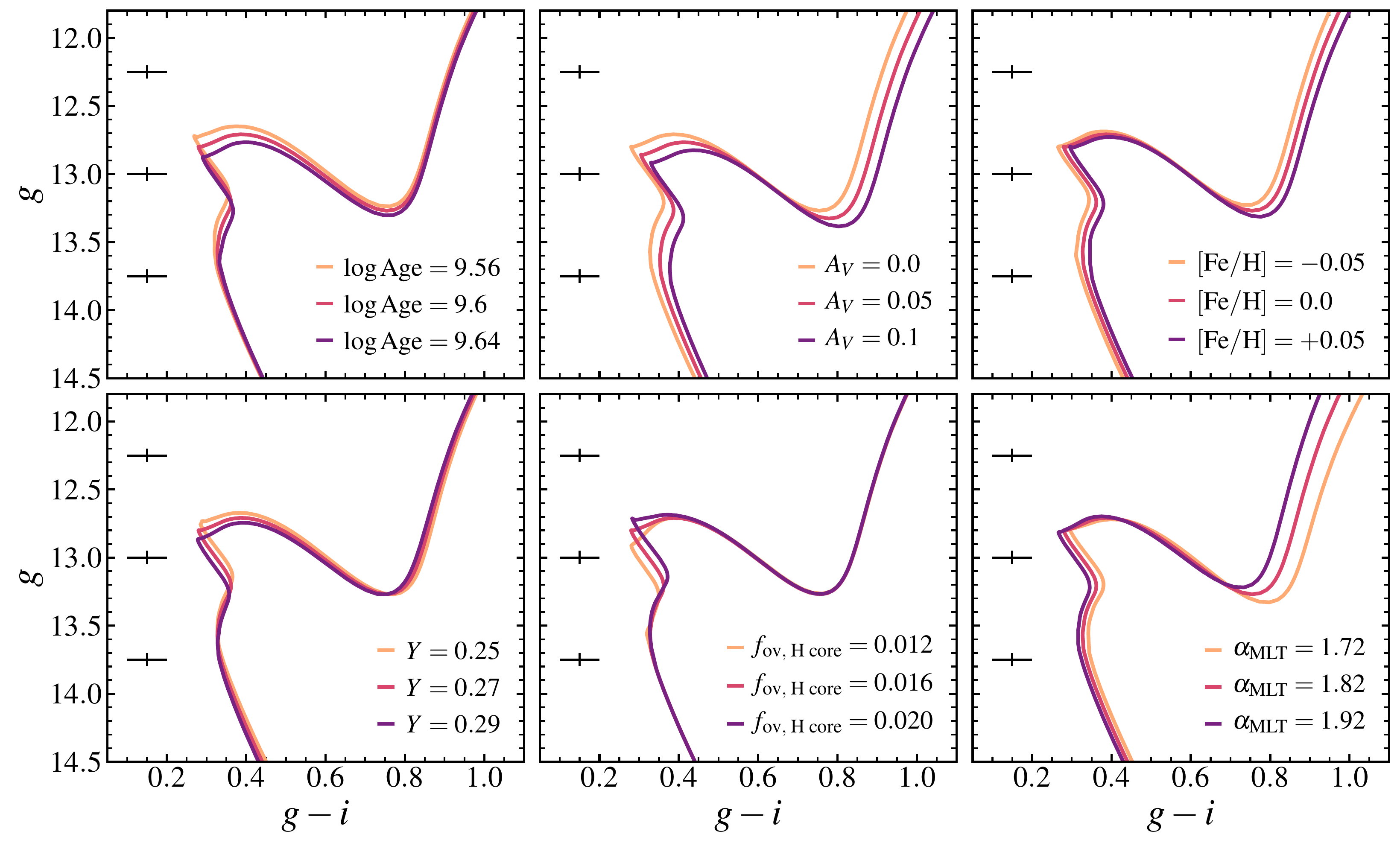}
\caption{Same as Figure~\ref{fig:ubv} except now showing {\it Pan-STARRS} $g$ vs. $g-i$.}
\label{fig:gig}
\end{figure*}

\begin{figure*}
\centering
\includegraphics[width=0.8\textwidth]{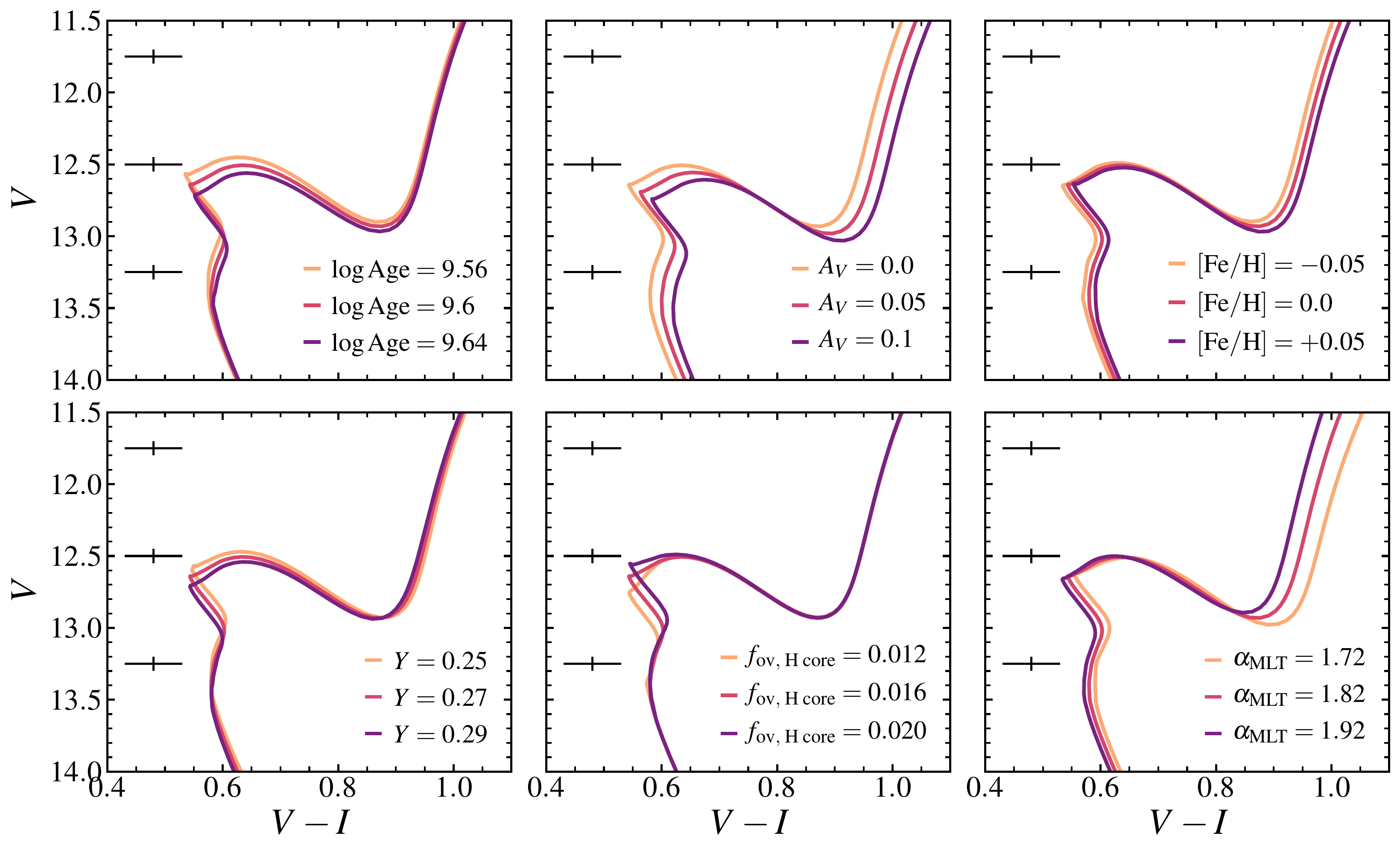}
\caption{Same as Figure~\ref{fig:ubv} except now showing $V$ vs. $V-I$.}
\label{fig:viv}
\end{figure*}

\begin{figure*}
\centering
\includegraphics[width=0.8\textwidth]{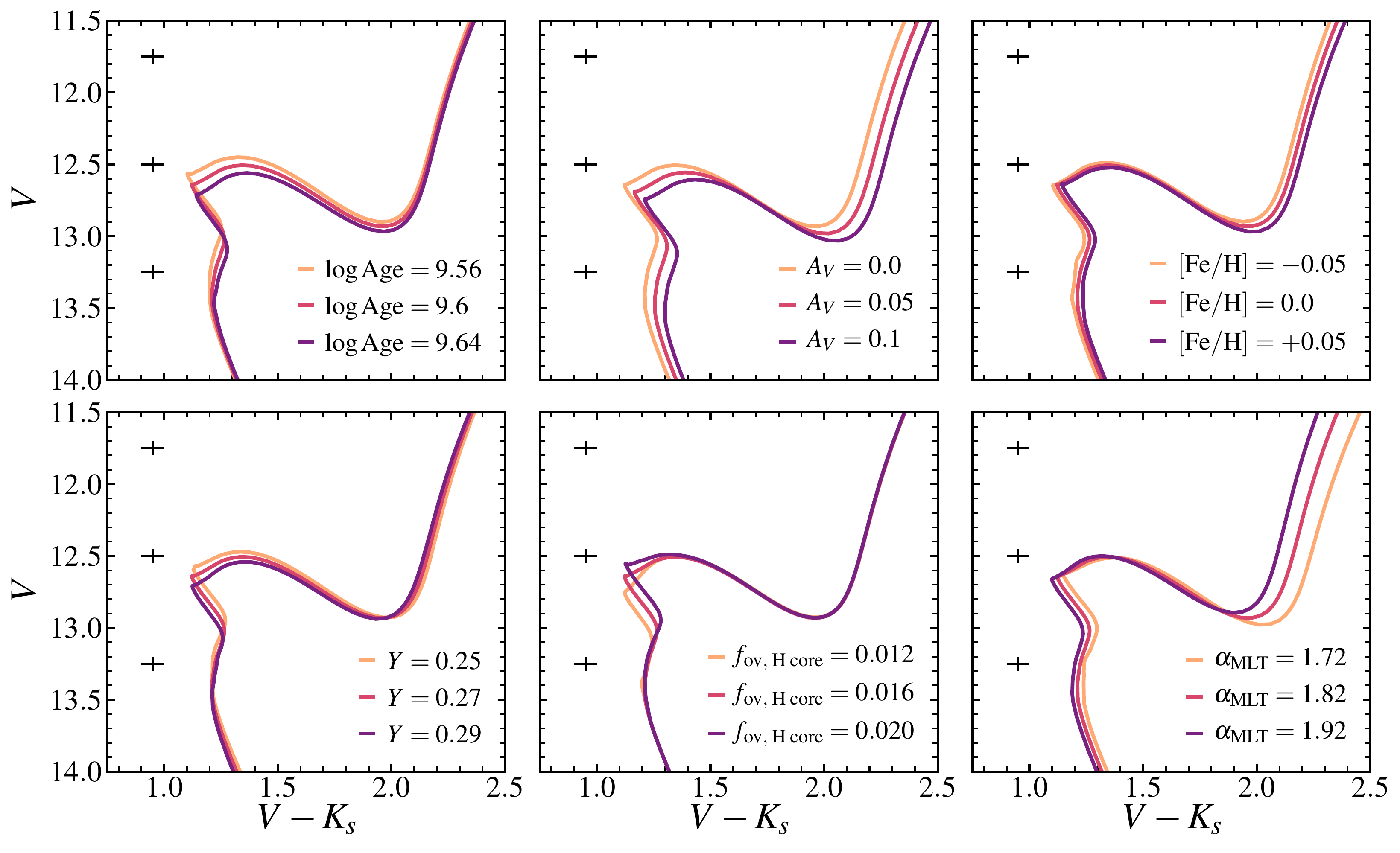}
\caption{Same as Figure~\ref{fig:ubv} except now showing $V$ vs. $V-K_s$}
\label{fig:vksv}
\end{figure*}

\begin{figure*}
\centering
\includegraphics[width=0.8\textwidth]{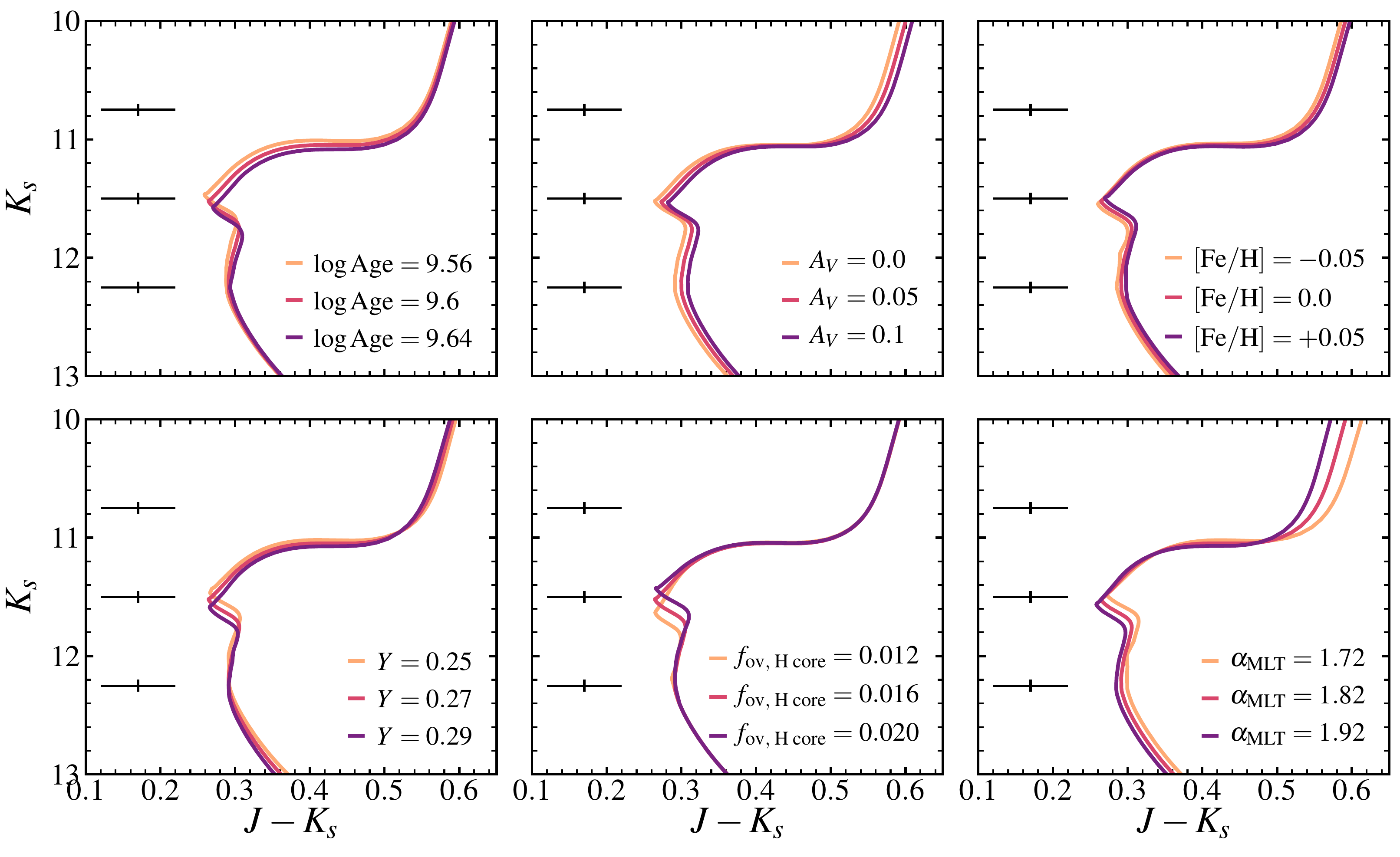}
\caption{Same as Figure~\ref{fig:ubv} except now showing $K_s$ vs. $J-K_s$}
\label{fig:jksks}
\end{figure*}

\begin{figure*}
\centering
\includegraphics[width=0.8\textwidth]{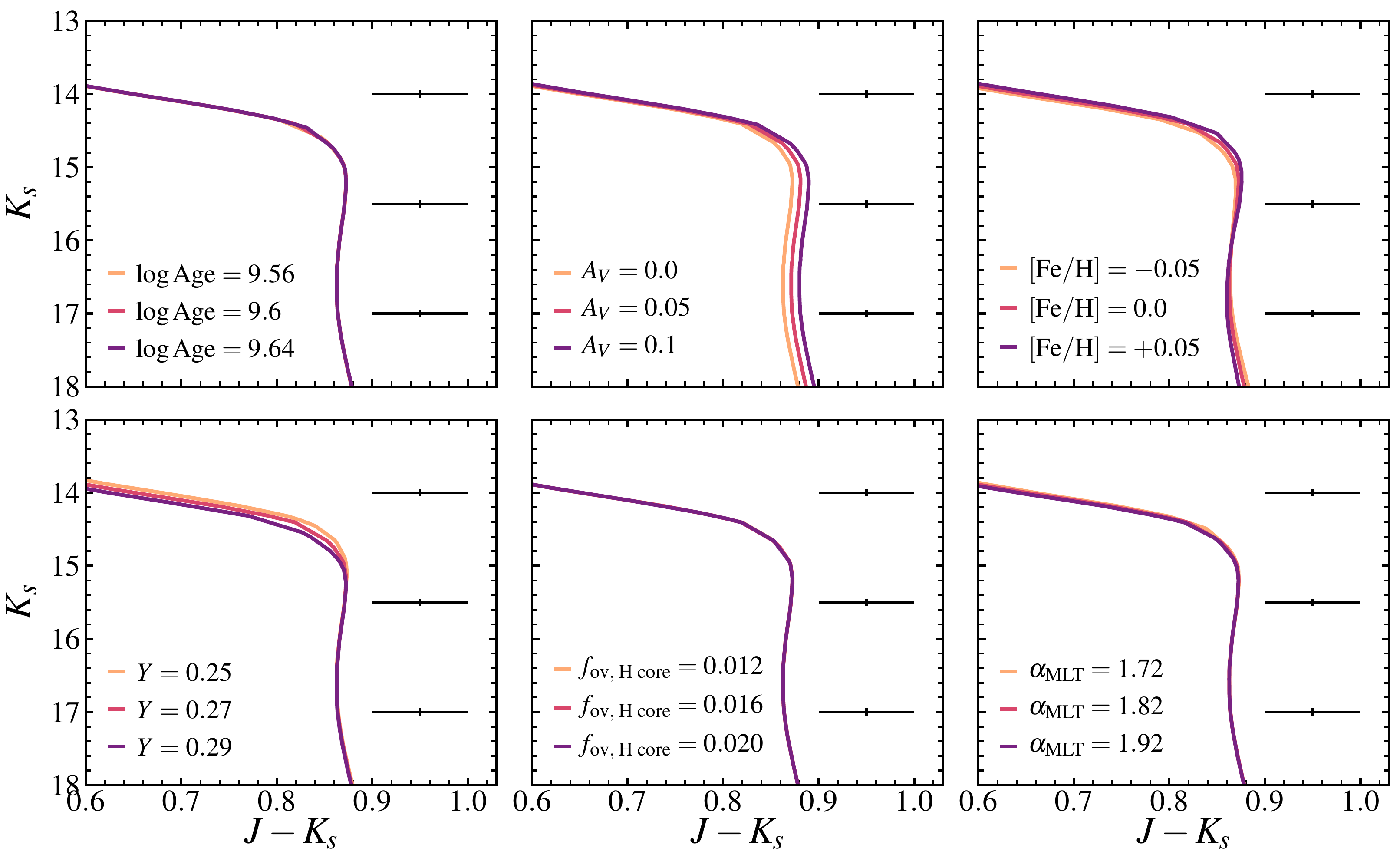}
\caption{Same as Figure~\ref{fig:jksks} except now zoomed in around the kink on the lower MS.}
\end{figure*}

\acknowledgments{}
We thank the anonymous referee for their feedback which greatly improved the quality of this manuscript. We also thank Lars Bildsten, Harshil Kamdar, Guillermo Torres, and Dan Weisz for helpful discussions and comments. Finally, we thank Bill Paxton and the rest of the MESA community who have made this project possible.

\bibliographystyle{apj}
\bibliography{bibtex.bib}

\end{document}